\input epsf

\documentstyle[12pt]{article}
\renewcommand{\baselinestretch}{1.3}
\newcommand{\asize}[1]{\renewcommand{\arraystretch}{#1}}

\newcommand{\be}{\begin{equation}}
\newcommand{\ee}{\end{equation}}
\newcommand{\dst}{\displaystyle}

\newcommand{\ssst}{\scriptscriptstyle}

\newcommand{\bal}{\begin{array}{l}}
\newcommand{\bac}{\begin{array}{c}}
\newcommand{\bacc}{\begin{array}{cc}}
\newcommand{\barcl}{\begin{array}{rcl}}
\newcommand{\eac}{\end{array}}

\renewcommand{\l}{\langle}
\newcommand{\r}{\rangle}

\newcommand{\tr}{{\rm tr}}

\newcommand{\Lbar}{{\overline{L}}}
\newcommand{\Nbar}{{\overline{N}}}
\newcommand{\mbar}{{\overline{m}}}
\newcommand{\zbar}{{\overline{z}}}

\newcommand{\alphabar}{{\overline{\alpha}}}

\newcommand{\rhobar}{{\overline{\rho}}}

\renewcommand{\d}{{{\partial}}}
\newcommand{\half}{\frac{1}{2}}
\newcommand{\atepi}{\frac{1}{4\pi \alpha ^{\prime}}}
\newcommand{\atepim}{\frac{-1}{4\pi \alpha ^{\prime}}}

\newcommand{\atepimi}{-4\pi \alpha ^{\prime}}

\newcommand{\p}{^{\prime}}

\newcommand{\ap}{\alpha ^{\prime}}
\newcommand{\ra}{\rightarrow}

\newcommand{\Is}{{\scriptscriptstyle I}}
\newcommand{\Js}{{\scriptscriptstyle J}}
\newcommand{\Ks}{{\scriptscriptstyle K}}
\newcommand{\Ls}{{\scriptscriptstyle L}}

\newcommand{\Hphys}{{\cal H}_{phys}}
\newcommand{\DD}{{\cal D}}
\newcommand{\JJ}{{\cal J}}
\newcommand{\LL}{{\cal L}}
\newcommand{\OO}{{\cal O}}
\newcommand{\PP}{{\cal P}}

\newcommand{\RR}{{\cal R}}

\renewcommand{\baselinestretch}{1.3}

\begin{document}

\thispagestyle{empty}
\setcounter{page}{0}

\begin{flushright}
PUPT-1338 \\ [-1mm]
RU-94-21\\ [-1mm]
hep-th/9402106\\ [-1mm]
February, 1994
\end{flushright}

\begin{center}
 ~ \\
{\Large Light-Cone Gauge Quantization of 2D Sigma Models} \\
 ~ \\
 ~ \\
{\large \sc ~Robert E. Rudd}$^{\dag}$ \\
 ~ \\
{\em Joseph Henry Laboratories} \\
{\em Princeton University} \\
{\em Princeton, New Jersey  08544} \\ [1mm]
and \\ [1mm]
{\em Dept. of Physics and Astronomy} \\
{\em Rutgers University} \\
{\em Piscataway, New Jersey~~08855-0849}
\end{center}

\vspace{.5in}
\centerline{\large ABSTRACT}
\centerline{~}

This work describes the formulation of the manifestly ghost-free
(spacetime) light-cone gauge for bosonic string theory with
non-trivial spacetime metric, antisymmetric tensor, dilaton and
tachyon fields.  The action is a general two-dimensional sigma model,
corresponding to a closed string theory with a second order action
in the Polyakov picture.  The spacetime fields must have a symmetry
generated by a null, covariantly constant spacetime vector in order
for the light-cone gauge to be accessible.  Also, the theory must be
Weyl invariant.  The conditions for Weyl invariance are computed within
the light-cone gauge, reproducing the usual beta functions.  The
calculation of the dilaton beta function and the critical dimension is
somewhat novel in this ghost-free theory.  Some exactly solvable
light-cone theories are discussed.

\vfill
\centerline{To Be Submitted to {\em Nuclear Physics B}}

\vspace{.3in}
\footnoterule
$^{\dag}$ \parbox[t]{6in}{
{\small E-mail: ~~rerudd@physics.rutgers.edu} }

\newpage

\section{Introduction}
\label{subsec-LCintro}

The light-cone gauge played an important role in the early
development of string theory.  It provided a way to consistently
quantize the Nambu-Goto string from first principles, giving a
manifestly unitary theory \cite{GGRT,NG}.
The bosonic string was found to suffer from
an anomaly in the Lorentz algebra unless the number of spacetime
dimensions is 26, and the Regge intercept is 1.  This was the first
definite problem encountered with string theory in less than the
critical dimension.  The operator formalism had required $D\le 26$ to be
ghost-free \cite{noghost}, and it had favored the critical dimension in
order to have proper factorization
of open string amplitudes \cite{factor}, but it was not clear that the
theory was inconsistent in fewer than 26 dimensions.

String interactions were incorporated into the light-cone
theory \cite{Mand1}, and the three string vertex was
found to be Lorentz invariant
exactly in the critical dimension \cite{Mand2}.  Scattering amplitudes were
calculated using Neumann function techniques.
The theory was then reformulated as
the first string field theory \cite{LCSFT}.
The Mandelstam diagrams of the first
quantized theory were cut into propagators and vertices, and the theory
was second quantized.  The Feynman rules for both the open, and
the closed bosonic string were derived.  Light-cone gauge was also
instrumental in the development of fermionic string theory
\cite{FLCSFT}.  The supersymmetric versions of
light-cone gauge quantum mechanics, first quantized field theory and
second quantized field theory have been developed as well \cite{SSLCSFT}.

Then the light-cone gauge was largely abandoned in favor of formulations
in which the target space is treated covariantly.  The advent of the
Polyakov picture revolutionized string theory \cite{Poly2}.  The critical
dimension
was identified with the vanishing of the Weyl anomaly.  Ghosts were
introduced to fix the intrinsic geometry of the worldsheet \cite{Poly2}.
The covariant formulations that followed---conformal field
theory \cite{CFT}, the operator formalism \cite{opform}
and covariant string field theory \cite{CSFT}---have
the advantage of not singling out any direction in spacetime.  Also,
there is a gauge symmetry, conformal invariance, that is very powerful
in analyzing the theory.
Despite the fact that the covariant formulations
retain more unphysical degrees of freedom, scattering amplitudes at the
zero- and one-loop order are calculated more easily than in light-cone
gauge.  This is because there is no complete operator formalism for the
light-cone gauge.  The tree level amplitudes are easy when $p^+=0$ for
all but two of the strings; otherwise, operator techniques cannot be
used since $e^{ip^+X^-}$ is not well-defined \cite{Mand1}.

The Nambu-Goto action had allowed for string propagation in curved
spacetime. A non-trivial spacetime metric could enter the action in
the Polyakov picture, too, but other spacetime fields could be added to
the action, as well.  The dilaton, in particular, was found to be vital
to renormalization on a curved worldsheet \cite{FT}.
Sigma model perturbation theory and background field techniques
were developed to determine the conditions for Weyl invariance
\cite{FT}-\cite{GSW2}.
The resulting beta functionals generalized the concept of the critical
dimension, requiring the spacetime background fields to satisfy
differential equations to assure Weyl invariance.  When these equations
are violated, the worldsheet theory is not scale invariant, so the
Liouville mode must be quantized as well, leading to non-critical
strings.

The development of the covariant formulations has been independent of
the light-cone gauge for the most part, after the invention of conformal
field theory.  The one striking exception is the covariant closed string
field theory.  The required non-polynomial action was elusive, and a
great deal of effort went into studying the relatively simple light-cone
gauge closed string field theory.  It only has three string vertices and
no contact terms.  It still provides one of the few
known triangulations of moduli space \cite{triang}.

The goal of this article is to quantize the general two-dimensional
bosonic sigma model in the light-cone gauge.  This amounts to using the
light-cone gauge to quantize closed string theory in the Polyakov
picture.\footnote{This is not to be confused with Polyakov's light-cone
gauge quantization of two-dimensional quantum gravity.}  The
worldsheet geometry plays a fundamental role, since the light-cone gauge
fixes the worldsheet reparameterization invariance.  This is done
without first going to the conformal gauge (as advocated by \cite{GSW}).
The light-cone gauge avoids propagating reparameterization ghosts.
It eliminates the negative norm modes, at the expense of manifest
worldsheet covariance.

The light-cone gauge is found to be accessible and
non-singular provided that all the spacetime fields
(the metric, antisymmetric tensor
field, dilaton and tachyon backgrounds) have a symmetry generated by a
null covariantly constant spacetime vector.  Also, the spacetime fields
must satisfy the usual Weyl invariance conditions (vanishing beta
functions) as in the conformal gauge---i.e.\ it must be a critical string
theory.  The way these conditions arise is somewhat novel, especially
the $(D-26)$ term of the dilaton beta function in this ghost-free
theory.  These conditions guarantee a consistent string quantum
mechanics.  They also seem to be enough to allow interactions and a
consistent string field theory, but we will not go through a complete
analysis.

Having recounted a few of the many wonderful successes of the covariant
formulations, one might well ask why the light-cone gauge should be
revived in the more general Polyakov framework.  Will it only provide a
cumbersome check of results from the covariant techniques?  There are two
applications of current interest which could benefit greatly from a
light-cone gauge analysis.  In both cases it is the fact that the
light-cone gauge explicitly retains only physical degrees of freedom
that makes it useful.
All of the unphysical degrees of freedom, like the Lorentz and
reparameterization ghosts, are removed by gauge fixing.

The first application is the plane-fronted gravitational waves, and
their stringy generalizations \cite{pp,comprop}.
Sigma models with these target spaces
have been shown to be Weyl invariant to all orders in sigma model
perturbation theory.  Perhaps they are exactly solvable.  Covariant
techniques could be used for the analysis.  They
are not intractable since the sigma model perturbation theory stops at
one loop in any calculation. Even so, the bookkeeping is unwieldy.
Light-cone gauge simplifies the calculations enormously, and the
generalized plane-fronted wave
background fields meet the requirements for light-cone gauge
quantization exactly.  In fact, Horowitz and Steif have already used
light-cone gauge quantum mechanics to study a string propagating through
a plane-fronted wave to lowest order in $\ap$ \cite{HorSt}.
Our work puts their analysis on a sound footing, proving
unitarity, for instance.  It also
prepares the way for a full string field theory treatment using the
relatively simple light-cone closed string field theory.

The second application is the class of sigma models with two-dimensional
target spaces coming from Liouville theory and non-critical strings
\cite{noncrit}.
These models of $c \le 1$ matter coupled to two-dimensional quantum
gravity were first solved by matrix models \cite{matrix}.
The continuum solutions are
less complete, and many mysteries remain.  One of their most striking
features is the presence of physical states at discrete values of the
momentum.  The naive light-cone Hilbert space would be trivial---nothing
but the tachyon---because there are no transverse dimensions.  It is
interesting to see why the usual light-cone gauge fails, and how it
might be generalized to include the special states.  This turns out to
be a subtle problem \cite{Smith}, and its solution will be presented
elsewhere \cite{c1gr}.

Having cited some possible applications, we are in a position to state
what we should demand of a light-cone quantization, since we might be
willing to forgo some of the usual properties of the light-cone gauge.
In order of increasing stringency, these properties are the elimination
of as many unphysical degrees of freedom as possible, the elimination of
Lorentz ghosts, and the absence of all ghosts so that the theory is
manifestly unitary.  In addition, we might demand that the theory have a
string field theory representation, perhaps even a simple one.  We
should also demand that we be able to check the consistency of the
theory from within the light-cone gauge.  Traditionally this is done by
computing the Lorentz anomaly, but it is a big restriction to demand
that spacetime have a Lorentz isometry between $X^+$ and two transverse
dimensions.  The formulation that we will develop has all of the
usual light-cone gauge properties (except the Lorentz isometry), but
we will point out where some of these restrictions could be eased in
future applications.

\section{Gauge Fixing}
\label{subsec-GF}

The generating functional $Z$ for bosonic closed string theory in the
Polyakov picture is (see \cite{GSW} for a review)
\be
\renewcommand{\arraystretch}{1.8}
\barcl
Z & = & {\dst \sum _{topologies}
\int [\DD g_{ab}] \, \DD X^{\mu} e^{iS[g_{ab},X]}}\\
S & = & {\dst \atepim \int d^2\sigma \, \sqrt{g}
\left\{ g ^{ab} G_{\mu \nu}\d _a X^{\mu} \d _b X^{\nu}
+\frac{\epsilon ^{ab}}{\sqrt{g}} B_{\mu \nu}\d _a X^{\mu} \d _b X^{\nu}
- \ap R^{(2)} \Phi + T
\right\} . }
\label{SD}
\eac
\renewcommand{\arraystretch}{1.4}
\ee
The action $S[g_{ab},X]$ is a functional of the worldsheet
metric $g_{ab}$ and the $D$ spacetime coordinate fields $X^{\mu}$.
These fields live on a worldsheet parameterized by the coordinates
$\sigma ^a$ where $\sigma ^0 = \tau$, $\sigma ^1 = \sigma$,
and $\sigma $ ranges from $0$ to $2\pi$.
Both the spacetime and the worldsheet have the
Lorentzian signature $(-,+,\cdots,+)$.
The spacetime fields consist of the metric $G_{\mu \nu}(X)$, an
antisymmetric tensor field $B_{\mu \nu}(X)$, the dilaton $\Phi(X)$
and the tachyon $T(X)$.  They are functionals of the fields $X^{\mu}$.
The tachyon would be absent from the corresponding supersymmetric model,
and it has been included primarily due to the important role it plays in
two dimensions where it is massless.
$\epsilon ^{ab}$ is the Levi-Civita antisymmetric tensor (density) such
that $\epsilon ^{01}=1$.  Note that spacetime indices are Greek,
worldsheet indices are Roman, and repeated indices are summed.
The action is usually Wick rotated to a Euclidean worldsheet where the
sum over topologies becomes a sum over the genera of compact Riemann
surfaces.  We will perform the Wick rotation before quantization in
section 3.

The integral over $g_{ab}$ denotes an integral over
worldsheet metrics modulo worldsheet diffeomorphisms.
The action is invariant under reparameterizations of the
world-sheet coordinates where the fields transform as
\be
\barcl
\delta X^{\mu} & = & \epsilon ^a \d _a X^{\mu}
+ \cdots \\
\delta g^{ab} & = & -(\nabla ^a \epsilon ^b + \nabla ^b \epsilon ^a )
= \epsilon ^c \d _c g^{ab} - g^{ac} \d _c \epsilon ^b
- g^{bc} \d _c \epsilon ^a \\
\delta \sqrt{g} & = & {\dst -\half \sqrt{g} \, g_{ab} \,
\delta g^{ab} = \sqrt{g} \, \nabla _a \epsilon ^a =
\d _a \left( \epsilon ^a \sqrt{g} \right) }
\eac
\label{reparams}
\ee
under the diffeomorphism $\sigma ^a \ra \sigma ^a + \epsilon ^a (\sigma
, \tau )$.  The $X^{\mu}$ transformation is typically very complicated
and not exactly calculable (with non-trivial background fields),
but it may be calculated to low orders in sigma model perturbation theory.
For our purposes it is enough to know that the correction term
transforms as a spacetime vector depending on $\Phi$ and $T$
classically and on all the background fields in the quantum theory.

For certain configurations of the space-time metric, antisymmetric tensor,
dilaton and tachyon backgrounds, the theory is also invariant under Weyl
scalings
\be
\barcl
{\dst \delta g^{ab} } & = & {\dst \Lambda (\sigma ^c) \, g^{ab} } .
\eac
\label{Weyl}
\ee
The background fields will be chosen such that the Weyl
anomaly cancels and the Liouville mode decouples in the quantum theory.
The Weyl mode does not decouple for generic background configurations.
In fact, the trace of the stress tensor (which generates Weyl
transformations) takes the form
\be
\sqrt{g} \, T^a _{\, a} =
\beta ^T (X) \, \sqrt{g} +
\beta ^{\Phi} (X) \, \sqrt{g} \, R^{(2)} +
\beta ^G_{\mu \nu} (X) \, \sqrt{g} \d _a X^{\mu} \d ^a X^{\nu} +
\beta ^B_{\mu \nu} (X) \, \epsilon ^{ab} \d _a X^{\mu} \d _b X^{\nu}
\label{Ttrace}
\ee
in conformal gauge.
The beta functionals $\beta ^G_{\mu \nu}, \, \beta ^B_{\mu \nu},
\, \beta ^\Phi$ and $\beta ^T$ must vanish
for the action to be Weyl invariant.  Actually, the beta functions are
equivalent to the equations of motion for the massless fields entering
the low-energy effective action, so they must vanish if the theory is to
be sensible.  This is not a
requirement put in by hand.  It is difficult to calculate the beta
functionals exactly for general actions, but they may be found using a
perturbation expansion in $\ap$.  The well-known result is
\be
\barcl
\beta^G_{\mu\nu}&=&{\dst R_{\mu\nu} +2\nabla _{\mu}\nabla _{\nu}\Phi
 -\frac{1}{4}H_{\mu}^{~\lambda \sigma} H_{\nu \lambda \sigma}
 -\d _{\mu}T\d _{\nu}T+\cdots }\\
\beta^{\Phi}&=&{\dst -\frac{\ap}{16\pi ^2}\left[ \frac{26-D}{3\ap} +
 R-4\d _{\mu}\Phi\d ^{\mu}\Phi+4\nabla ^2\Phi -\frac{1}{12}
 H^2 -\d _{\mu}T\d ^{\mu}T+2T^2+\cdots \right]}\\
\beta^B_{\mu\nu}&=&{\dst \nabla _{\lambda} H^{\lambda}_{\mu \nu}
 -2 (\nabla _{\lambda}\Phi ) H^{\lambda}_{~\mu \nu} + \cdots }\\
\beta^T&=&{\dst -2\nabla ^2 T+4G^{\mu\nu}\d _{\mu}\Phi\d _{\nu}T-4T
+\cdots }
\eac
\label{bfns}
\ee
where
the field strength for the antisymmetric tensor field is given by
$H_{\alpha \beta \gamma } =  \d _{\alpha}B_{\beta \gamma }
+ \d _{\beta }B_{\gamma \alpha} + \d _{\gamma }B_{\alpha \beta }$
\cite{CFMP}.
We will consider sigma models for which these beta functionals are zero,
for now, in order to formulate the general light-cone gauge
quantization.  Once the formalism is established, we will return to the
question of the Weyl anomaly within the light-cone gauge.
If other (non-critical) backgrounds are used as the
starting point for quantization, additional degrees of freedom (such as
the Liouville mode) enter during quantization such that the equations
are solved in the end.  The Liouville mode couples through the Weyl
anomaly, for instance.  This is somewhat awkward to treat in light-cone
gauge, so we will
postpone this discussion.  For now, we assume that we are dealing with a
general critical string theory.

The basic idea with the light-cone gauge is to use reparameterization
invariance to gauge away the oscillator contribution to one of the
$X^{\mu}$ coordinates---in particular a null (light-cone) coordinate.
In fact, the reparameterization group is large enough to do this and
at the same time to fix a conformally flat world-sheet metric.
The conformal factor may then be set to unity in a Weyl invariant
theory.  The
oscillators may be eliminated from any one of the coordinates, but when
a null coordinate is chosen, the other null coordinate  becomes an
auxiliary field.  It may be expressed in terms of
the transverse coordinates through the constraints,
so that only the physical degrees of
freedom enter the gauge fixed theory.  There are no ghosts.

This section will develop the formalism classically.  The approach we will
take is somewhat unconventional, but it is appropriate for quantizing
the Polyakov theory in the second order formalism.  It only works for
critical string theories, where the beta functionals vanish.  The
standard light-cone gauge
approaches to quantizing the $D=26$ bosonic string
include quantizing the Nambu-Goto action \cite{GGRT,Mand1}, quantizing
the second order action by choosing the light-cone gauge subsequent to
choosing the conformal gauge \cite{GSW} and quantizing using hybrid
light-cone/conformal gauges \cite{Tzani}.  These approaches are either
inappropriate for the quantization of an action with a dilaton term
(such as Nambu-Goto) or they have reparameterization ghosts.
Another option is to quantize the first order Polyakov action
\cite{Siegel}, but this is inconvenient for the discussion of general
sigma model backgrounds.  So we will quantize the
Polyakov action in the second order formalism.

Consider the variation of the action (\ref{SD}) with respect to the fields
$X^{\mu}$ and $g^{\mu \nu}$.  This yields the classical field equations
\be
\barcl
{\dst 0 = \atepimi \frac{\delta S}{\delta X^{\mu}} } &  = & {\dst
-2 G_{\mu \nu} \left( \Delta X^{\nu} +
\Gamma ^{\nu} _{~\alpha \beta} g^{ab}
\d _a X^{\alpha} \d _b X^{\beta} \right) }\\
& & ~~~~~~~~{\dst
+ \frac{\epsilon ^{ab}}{\sqrt{g}} H_{\mu \alpha \beta}
\d _a X^{\alpha} \d _b X^{\beta}
- \ap R^{(2)} \, \d _{\mu} \Phi + \d _{\mu} T } \\
{\dst 0 =\frac{\atepimi}{\sqrt{g}}\frac{\delta S}{\delta g^{ab}} } & =
& {\dst G_{\mu \nu}\d _a X^{\mu} \d _b X^{\nu} - \ap R_{ab} ^{(2)} \Phi
+ \ap (\nabla _a \nabla _b X^\mu ) \d _{\mu} \Phi
+ \ap (\d _a X^\mu \d _b X^\nu ) \d _{\mu} \d _{\nu} \Phi
} \\
& & ~~~~~~~~{\dst
-\half g_{ab} \left\{ g ^{cd} G_{\mu \nu}\d _c X^{\mu} \d _d X^{\nu}
- \ap R ^{(2)} \Phi + 2\ap \nabla ^c \nabla _c \Phi + T \right\}
 }
\eac
\label{classEoM}
\ee
where the Laplacian is given by $\Delta = \frac{1}{\sqrt{g}} \d _a
\sqrt{g} g^{ab} \d _b$.
The variation with respect to the worldsheet metric is the stress-energy
tensor, $T_{ab}$.
Because string theory is Weyl invariant, it is useful to
decompose the metric into a Weyl part, $e^{\phi}$,
and a unit determinant part, $\gamma _{ab}$.
That is,
\be
g^{ab} = e^{-\phi (\sigma ^c)} \gamma ^{ab} ~~~~~{\rm with~~}
\det \gamma ^{ab} =-1.
\ee
Note that $\gamma ^{ab}= \sqrt{g} g^{ab}$ has two degrees of freedom.
The covariant metric on the space of $\gamma$'s is given by
\be
| \delta \gamma ^{ab} | ^2 = \int d^2 \sigma \, \gamma _{ac}
\gamma _{bd} \delta \gamma ^{ab} \delta \gamma ^{cd}.
\label{metmeas}
\ee
Rewriting the action with this metric decomposition, we find
\be
\renewcommand{\arraystretch}{1.7}
\barcl
Z & = & {\dst \sum _{topologies}
\int [\DD \phi ] \, \DD \gamma _{ab} \, \DD X e^{iS[g_{ab},X]}}\\
S & = & {\dst \atepim \int d^2\sigma \,
\left\{ \gamma ^{ab} G_{\mu \nu}\d _a X^{\mu} \d _b X^{\nu}
+\epsilon ^{ab} B_{\mu \nu}\d _a X^{\mu} \d _b X^{\nu}
- \ap  (-\Delta ^{(\gamma)} \phi + R^{(\gamma )} ) \Phi
+ e^{\phi} T \right\} }
\label{SDtwo}
\eac
\renewcommand{\arraystretch}{1.4}
\ee
where the integral over $\phi$ is trivial in the full quantum theory due
to the required Weyl invariance, and the $\gamma$ integral runs over the
two independent components.  The variation of the action with respect to
the metric (i.e.\ the stress tensor) may be reexpressed as
\be
\barcl
0 & = & {\dst \atepimi \frac{\delta S}{\delta \gamma ^{ab}}   =
G_{\mu \nu}\d _a X^{\mu} \d _b X^{\nu} - \ap R _{ab}^{(2)} \Phi
+ \ap (\nabla _a \nabla _b X^\mu ) \d _{\mu} \Phi
+ \ap (\d _a X^\mu \d _b X^\nu ) \d _{\mu} \d _{\nu} \Phi
} \\
& & ~~~~~~~~~~ {\dst - \half \gamma _{ab} \left[ \left( G_{\mu \nu}
+ \ap \d _{\mu} \d _{\nu} \Phi \right) \gamma ^{cd}
\d _c X^{\mu} \d _d X^{\nu} - \ap R^{(2)} \Phi
+ \ap (\nabla ^2 X^{\mu} ) \d _{\mu} \Phi \right] } \\ [2mm]
0 & = & {\dst \left. \atepimi \frac{\delta S}{\delta \phi}
\right|_{classical} =
\ap \Delta ^{(\gamma)} \Phi + e^{\phi} T }
\eac
\label{gammaEoM}
\ee
The Weyl anomaly cancels the classical variation of the action with
respect to $\phi$ in the quantum theory.
It will be convenient to define the $\gamma$ stress tensor,
\be
T^{(\gamma)} _{ab} = \atepimi \frac{\delta S}{\delta \gamma ^{ab}} =
T_{ab} - \half \gamma _{ab} ( \ap \Delta ^{(\gamma)} \Phi + e^{\phi} T).
\ee
It is
traceless since $\det \gamma ^{ab} = -1$.  Note that
$T_{01} = T^{(\gamma )} _{01}$ in gauges with $\gamma ^{01} =0$.

The key to the light-cone gauge is that when the two fields $X^+$
and $\gamma ^{00}$ are fixed, the resulting constraints insure that
two other fields, $X^-$ and $\gamma ^{01}$, are auxiliary.  In fact,
these fields enter the constraints linearly, so the constraints may be
used to solve for them in terms of the transverse fields.  This
eliminates the maximum number of degrees of freedom explicitly.  In
the process the worldsheet metric is fixed to be flat, as required
ultimately in order to express scattering amplitudes in terms of the
usual Mandelstam diagrams for light-cone gauge.  We will examine how
this works in detail.

The main point of this section is to show under what conditions the
light-cone gauge is non-singular.  The gauge is not accessible with
generic target space fields.  First, the gauge conditions must solve the
classical equations of motion.  Then, the resulting constraints
must be solvable.  This is what prohibits the light-cone gauge for
a flat $D<26$ target space (upon quantization).  Finally,
it should be possible to solve certain constraints as operator
equations for the fields $X^-$ and $\gamma ^{01}$.
The gauge fixing does not require Faddeev-Popov ghosts, so this
reduces the the theory down to the (off-shell) physical degrees of
freedom.  These criteria are
met for a large class of target space fields, as shown in section 4.

\subsection{Fixing $X^+$ and $\gamma ^{00}$}

Consider the following gauge choice:
\be
\barcl
X^{+}  & = & p^+ \tau + x^+_0 \\
\gamma ^{00} & = & -1
\eac
\label{gengauge}
\ee
where
\be
X^{\pm} = \frac{1}{\sqrt{2}} (X^1 \pm X^0).
\ee
According to equation (\ref{reparams}), infinitesimal
reparameterizations about this gauge choice give
\be
\barcl
\delta X^{+} & = & \epsilon ^a \d _a (p^+ \tau + x^+_0) + \cdots
 = p^+\epsilon ^0 + \cdots \\
\delta \gamma ^{00} & = & {\dst
\d _c \left[ \epsilon ^c (-1)\right] - 2 \gamma ^{0c} \d _c \epsilon ^0 }
= \d _0 \epsilon ^0 - \d _1 \epsilon ^1
\eac
\label{gvars}
\ee
when $\sigma ^a \ra \sigma ^a + \epsilon ^a (\sigma , \tau )$.
The additional terms in $\delta X^+$ generically
lead to ghosts.  They transform as the $+$ component of a spacetime
vector depending on $\Phi$ or $T$, classically.  After quantization and
renormalization, these terms also depend on $G_{\mu \nu}$ and
$B_{\mu \nu}$.  A condition sufficient to insure the absence of
Faddeev-Popov ghosts is that all such vectors vanish;
that is, $\d _- \Phi = \d _- T = \d _- G_{\mu \nu} = \d _- B_{\mu \nu} =
0$, and $G_{--} = G_{- i} = B_{- \nu} = 0$.
This condition arises from other considerations as well,
as we will see below.
Setting these terms to zero, the Faddeev-Popov determinant is
\be
\Delta _{FP} = \left|
\bacc
p^+ & 0 \\
\d _0 & - \d _1
\eac \right| .
\label{FPdet}
\ee
We have used the fact that $\gamma^{01}$ will be fixed to zero.
Since the relevant part (the diagonal) of the determinant
does not depend on time derivatives, it just contributes a
constant factor to the measure.
The corresponding ghosts do not propagate.
At the level of quantum mechanics,
the determinant could be absorbed into the overall normalization of
the path integral, except for the dependence on $p^+$.  It is odd that
the physical parameter $p^+$ should enter the normalization, which is
independent of the dynamics.  It turns out that
this $p^+$ dependence cancels that coming from the
eventual elimination of $X^-$.  In fact, we will see that rescaling the
worldsheet by $p^+$ ($\sigma \ra \sigma/p^+ , \tau \ra \tau /p^+$)
makes both determinants equal to unity.
This is especially important, since $p^+$ is not a Lorentz scalar.
When the spacetime metric has a global Lorentz isometry, it may be used
to check scattering amplitudes and to detect the occurrence of anomalies.
Lorentz invariance would be spoiled by $p^+$ in the normalization of
the path integral.

The gauge choice (\ref{gengauge}) does not completely fix the
reparameterization invariance, since the transformation
$\sigma ^1 \ra \sigma ^1 + \epsilon ^1 (\tau )$ leaves the gauge
condition unchanged.  The residual gauge invariance will be used to
eliminate $\gamma ^{01}$.  Then the gauge
(\ref{gengauge})
is accessible and unique on any Mandelstam worldsheet with at least
one vertex (The two point amplitude at tree level with its
cylindrical worldsheet has the residual invariance $\sigma \ra \sigma +
const.$ and $\tau \ra \tau + const.$).

\subsection{The Auxiliary Fields $X^-$ and $\gamma ^{01}$}

The gauge choice should also constrain the other null
coordinate for it to be a useful light-cone gauge.
Both $X^-$ and $\gamma ^{01}$ turn out to be auxiliary fields.
Consider the variation of the gauge fixed action with respect
to $\gamma ^{01}$:
\be
0 = \atepimi \frac{\delta S}{\delta \gamma ^{01}} =
G_{\mu \nu}\d _0 X^{\mu} \d _1 X^{\nu}
+ \ap (\nabla _0 \nabla _1 X^\mu ) \d _{\mu} \Phi
+ \ap (\d _0 X^\mu \d _1 X^\nu ) \d _{\mu} \d _{\nu} \Phi  .
\label{gzeroEoM}
\ee
This is a constraint since $\gamma ^{01}$ is auxiliary.
We would like to solve it for $X^-$, the other light-cone
coordinate, expressing it in terms of the transverse coordinates and
$p^+$.  Of course, $X^-$ is an operator, and the only known way to have
a sensible algebra is if it is expressed as a (differential) polynomial in
the other fields.  It cannot be realized as the square-root of the
transverse stress-tensor, for example.  So equation (\ref{gzeroEoM})
must be linear in $X^-$.\footnote{The only caveat is that the equation
might factorize.  If the dilaton is constant, then the spacetime metric
could be rescaled by an $X^-$ dependent conformal factor, $G_{\mu \nu} =
\Lambda (X^- ) \tilde{G} _{\mu \nu}$.  It would
factor out of (\ref{gzeroEoM}).  This possibility is ruled out by
requiring $\gamma ^{01} =0$.} Then
$G_{--}=0$, $\d _- G_{-\mu} = 0$, $\d _- \d _- G_{\mu \nu}=0$ and
$\d _- \d _- \Phi =0$.  Also, $G_{-i}=0$, since $\d _a
X^i$ cannot appear in the coefficient of $X^-$ if it is to be inverted.
Similarly, $G_{+-}$ must be constant, which will be chosen to be 1 by
rescaling $X^-$.  Now equation (\ref{gzeroEoM}) becomes
\be
\barcl
0 & = & {\dst p^+ \d _1 X^- +
G_{+i}p^{+} \d _1 X^{i} +
G_{ij}\d _0 X^{i} \d _1 X^{j}
+ \ap \left\{ (\nabla _0 \nabla _1 X^- ) \d _{-} \Phi  \right. }
\\
 & & ~~~~~ {\dst  \left.
+ (\d _0 X^- \d _1 X^i ) \d _{i} \d _{-} \Phi
+ (\d _1 X^- \d _0 X^i ) \d _{i} \d _{-} \Phi
+ (p^+ \d _1 X^- ) \d _{+} \d _{-} \Phi   \right\} } \\
 & & ~~~~~ {\dst + \ap (\nabla _0 \nabla _1 X^i ) \d _{i} \Phi
+ \ap (p^+ \d _1 X^i ) \d _{i} \d _{+} \Phi
+ \ap (\d _0 X^i \d _1 X^j ) \d _{i} \d _{j} \Phi  }
\eac
\label{varg}
\ee
Note that only a few terms coming from the dilaton contain $\tau$
derivatives of $X^-$.  If we require additionally that $\d _- \Phi
=0$, then the terms in the braces in
(\ref{varg}) vanish, and $X^-$ is an auxiliary field.
$X^-$ is expressed in terms of the transverse stress tensor as
\be
X^- = - \int ^\sigma d\sigma \p \,
\left\{ \frac{1}{p^+} T^{tr}_{01} (\sigma \p ,\tau)
+ G_{+i} \d _1 X^i + \ap \d _1 \d _+ \Phi \right\} + x^- (\tau )
\label{Xminus}
\ee
where the $\sigma$-independent piece $x^- (\tau)$ is undetermined.
It cannot be eliminated without using constraints involving time
derivatives.  The trivial integral over the dilaton term has not been
done as a reminder that the $\sigma$ independent part is missing.
Eliminating $X^-$ produces the determinant $ (\det \p
p^+ \d _1)^{-1}$.  The $p^+$ dependence cancels that in the Faddeev-Popov
determinant, and the remainder of the determinant is a constant which
may be absorbed into the normalization of the path integral.

Thus, we arrive at a simple expression if we impose
two sets of conditions on the background fields.  Since it may be
possible to relax these conditions, we will restate them.
The first set of conditions insures that $X^-$ has an explicit
representation on the transverse Fock space.  They are
\be
\barcl
G_{-\mu} & = & 0 ~~~{\rm except} ~~~ G_{-+} = 1 \\
\d ^2 _- G _{\mu \nu} & = & 0 \\
\d ^2_- \Phi & = & 0.
\eac
\ee
(In terms of the inverse metric the conditions are $G^{+\mu} = 0$ except
$G^{+-} =1$.)
The second set of conditions guarantee that $X^-$ is an auxiliary
field, so that it may be eliminated by its equations of motion without
producing a non-trivial Jacobian.  These conditions are
\be
\barcl
G_{--} & = & 0 \\
G_{-i} & = & 0 \\
\d _- \Phi &= & 0.
\eac
\ee
Once the first set of restrictions is imposed, the second set
only forbids a dilaton linear in $X^-$.\footnote{The linear dilaton is
important for $c\le 1$ Liouville theory.  Evidently, the usual
light-cone gauge will not suffice, but a generalized light-cone gauge is
possible \cite{c1gr}.}  The complete set of
restrictions is sufficient to prohibit the
Faddeev-Popov ghosts discussed deriving (\ref{FPdet}).

As an example of how these restrictions may be circumvented,
consider choosing the light-cone gauge subsequent to fixing the conformal
gauge in the usual D=26 bosonic string, as has been advocated by many
authors \cite{GSW}.  The 26 dimensional
spacetime with $G_{\mu \nu} = \eta _{\mu
\nu}$ and $B_{\mu \nu}=\Phi=T=0$ certainly meets the requirements for a
light-cone quantization given above, but the conformal gauge approach
is not ghost-free.  Gauge
fixing produces the non-trivial Faddeev-Popov determinants and ghosts
familiar from the usual conformal gauge treatment, along with additional
determinants and Jacobians.  All of these factors cancel in the end to
give a trivial measure, as required.  There is no complete discussion of
this cancellation in the literature, but it must occur since the gauge
fixed action is identical to the one we find, up to the measure.  It
would be interesting to see how this works.  Most of
the no-ghost statements in
light-cone gauge rely on the implicit equivalence with the Nambu-Goto
light-cone string, where the no-ghost theorem is more straight-forward
and well established.  While this example
does not violate the restrictions placed on the
background fields above, it does show how there may be cancellations
among the various measure factors which we have not considered.
Such cancellations can be very complicated, as they are in the conformal
gauge example.

It remains to show that the other $\gamma ^{ab}$ degree of freedom may
be eliminated using the constraints.
Since $X^-$ has been eliminated as an auxiliary field, varying the
action with respect to it yields a constraint,
\be
\barcl
{\dst 0 = \atepimi \frac{\delta S}{\delta X^-} } & = & {\dst
-2 G_{- \nu} \left( \Delta X^{\nu} +
\Gamma ^{\nu} _{~\alpha \beta} g^{ab}
\d _a X^{\alpha} \d _b X^{\beta} \right) }\\
& & ~~~~~~~~{\dst
+ \frac{\epsilon ^{ab}}{\sqrt{g}} H_{-\alpha \beta}
\d _a X^{\alpha} \d _b X^{\beta}
- \ap R^{(2)} \, \d _{-} \Phi + \d _{-} T. }
\eac
\label{xminusEoM}
\ee
In order to solve this constraint for $\gamma ^{01}$, all of the
dependence on the $X^i$ fields must vanish.  There is no cancellation
among the terms for general field configurations, so
\be
\Gamma _{-\alpha \beta} = H_{-\alpha \beta} = \d _- \Phi = \d _- T = 0.
\label{gsolvable}
\ee
Note that $\d _- \Phi =0$, as required above for $X^-$ to be an
auxiliary field.  Also, using the previous restrictions
on the metric, the vanishing Christoffel symbol becomes
\be
0 = \Gamma _{-\alpha \beta} = -\half \d _- G_{\alpha \beta} .
\ee
This also rules out the $X^-$ dependent conformal factor mentioned in
footnote 2.
The only field that can depend on $X^-$ is $B_{\mu \nu}$.  In fact,
the antisymmetric tensor possesses a gauge invariance that may be used
to eliminate the $X^-$ dependence from it, too.  The theory is invariant
under
\be
B_{\mu \nu} \ra B_{\mu \nu} + \d _{\mu} \Lambda _{\nu} -
\d _{\nu} \Lambda _{\mu}
\label{Bgauge}
\ee
where $\Lambda _{\mu} (X)$ is any functional of $X^{\mu}$.  Consider the
gauge transformation given by
\be
\barcl
{\dst \Lambda _{\alpha}} & = &
{\dst - \int ^{X^-} dX^{-\, \prime} B_{- \alpha} (X^+,X^{-\, \prime},
X^i) } \\
{\dst \Lambda _{-}} & = & 0
\eac
\ee
Then $B_{- \mu} \ra B_{- \mu} + \d _{-} \Lambda _{\mu} -
\d _{\mu} \Lambda _{-} = 0$.  The fields still satisfy
(\ref{gsolvable}), so $0=H_{-\alpha \beta}= \d _- B_{\alpha \beta} +
\d _{\alpha} B_{\beta -} + \d _{\beta} B_{- \alpha} =
\d _- B_{\alpha \beta}$.  Evidently, all of the spacetime fields are
independent of $X^-$ in this gauge.

\subsection{The Geometry of Light-Cone Sigma Models}

Horowitz and Steif \cite{HorSt} have investigated the conditions
necessary to impose the light-cone gauge subsequent to the conformal
gauge in a theory essentially with $\Phi = T = 0$ (at tree level in
sigma model perturbation theory).  They point out
that the conditions may be phrased geometrically:  the spacetime metric
must admit a covariantly constant null vector.  A generalization of
this condition
is necessary to get all the benefits of a light-cone gauge theory
in the general sigma models we have considered, as well.  The field
$K_{\mu} \equiv \d _{\mu} X^+$ is a
covariantly constant null vector.  It trivially satisfies the Killing
equation $\nabla _{(\alpha } K_{\beta )} =0$.  In fact, $K_{\mu}$ not
only generates an isometry in $G_{\mu \nu}$, but also symmetries in the
other fields
\be
\LL _K G_{\mu \nu} = \LL _K B_{\mu \nu} = \LL _K \Phi = \LL _K T = 0,
\label{Xminussym}
\ee
where $\LL _K$ is the Lie derivative in the $K_{\mu}$ direction.
Note that this is
an invariant statement, not restricted to the special coordinates suited
for the light-cone gauge.  Thus,
to have a full light-cone quantization, all of the spacetime fields must
have a symmetry generated by a single, covariantly constant null
vector (in the $X^-$ free gauge for $B_{\mu \nu}$).

There are spacetime field configurations which satisfy both the
background field equations for Weyl invariance (\ref{bfns})
and the light-cone gauge quantization conditions (\ref{Xminussym}).
This is exactly the kind of configuration
studied in ``Compactification Propagation'' \cite{comprop}.
For example, consider the metric and dilaton given by
\be
\barcl
ds^{2} & = & {\dst dX^+\, dX^- + \sum _{i}~[2\pi R_{i}(X^+)]^{2}
(dx^{i})^{2} } \\ [2mm]
\Phi (X^+) & = & {\dst \half \int ^{X^+} \int \sum _{i}
\frac{R_{i}^{\prime \prime}} {R_{i}}}
\eac
\label{cp_ex}
\ee
where $R_{i}(X^+)$ for $i=1,\cdots,24$ is an arbitrary function of
$X^+$, and the primes denote differentiation with respect to $X^+$.
These fields certainly satisfy the light-cone gauge requirements, since
they are independent of $X^-$.  They also satisfy the background field
equations to all orders in $\ap$, as shown in the reference
\cite{comprop}.
Since the fields meet both requirements, the corresponding sigma model
may be quantized in the light-cone gauge.
This is just one of a large class of light-cone sigma models.
We will consider more examples below.

\subsection{Additional Constraints and Weyl Invariance}

We now consider only those backgrounds meeting the light-cone gauge
requirements.  Then the equation (\ref{xminusEoM}) reduces to
\be
\Delta X^+ = e^{\phi} \d _1 \gamma ^{10} p^+ =0
\label{gzero}
\ee
so $\gamma ^{10}$ must be independent of $\sigma$.
Under an infinitesimal reparameterization about $\gamma ^{10}=0,
\gamma ^{00}=-1$ and $\gamma ^{11}=1$, the variation of $\gamma ^{10}$
is
\be
\delta \gamma ^{10} =
\d _c \left[ \epsilon ^c (\gamma ^{10})\right]
- \gamma ^{0c} \d _c \epsilon ^1
- \gamma ^{1c} \d _c \epsilon ^0
= \d _0 \epsilon ^1 - \d _1 \epsilon ^0.
\label{gzerovar}
\ee
We can now use the residual reparameterization invariance
$\sigma ^1 \ra \sigma ^1 + \epsilon ^1 (\tau )$
to set $\gamma ^{10} = 0$ at
one value of $\sigma$, say $\sigma _0$.  Then the equation of motion
guarantees that it vanishes everywhere, giving the Jacobian
$[p^+ \det \d _1]^{-1}$.  This fixes two degrees of
freedom of the worldsheet metric, leaving only the Weyl mode.
It decouples from the quantum theory according to our ansatz.

It should be noted that while $\gamma ^{10}$ is an auxiliary field,
it is partially gauge fixed using (\ref{gzerovar}) which contains a time
derivative.  Fortunately, the resulting  determinant is trivial and does
not require ghosts.  The would-be ghost action is given by
\be
S^{ghost}_{\gamma ^{10}} = \int d\tau b \, \d _0 c
\ee
where $b$ and $c$ are the ghosts.  They only depend on $\tau$,
not $\sigma$.  The ghosts do not interact, and they do not depend on
the worldsheet geometry, so they just contribute to the overall
normalization of the path integral.  They are irrelevant.
This is important, because flat Minkowski space is a
special case of the general backgrounds studied here, and
the measure is known to be flat in that case.

The variation of the action with respect to the gauge fixed field
$\gamma ^{00}$ gives the Virasoro constraint $T^{(\gamma )}_{00} =0$.
The zero mode ($\sigma$--independent) part of this constraint is just the
mass-shell condition.  The non-zero mode part is less obvious.  Consider
\be
\asize{1.8}
\barcl
{\dst 0 = \left. \frac{\delta S}{\delta \gamma ^{00}} \right| _{nzm} }
 & = & {\dst \left\{ T_{00}
- \half g _{00} \tr T^{(g)} \right\} _{nzm} } \\
 & = & {\dst \left\{ T_{11} - \half \sqrt{g} \, \tr T^{(g)}\right\} _{nzm}}\\
 & = & {\dst \left\{ \int ^{\sigma} \d _0 T_{01}
- \half \sqrt{g} \, \tr T^{(g)} \right\} _{nzm} } \\
 & = & {\dst \left. - \half \sqrt{g} \, \tr T^{(g)} \right| _{nzm} }
\eac
\asize{1.4}
\label{gzzEoM}
\ee
where $\tr T^{(g)}= g^{ab} T^{(g)}_{ab}$, and we have used the stress
tensor conservation equation $\nabla ^a T_{ab}\sim 0$.
Classically the stress tensor conservation equation reads
\be
\nabla ^a T_{ab} = 2 \pi \ap \d _b X^{\mu}
\frac{\delta S}{\delta X^{\mu}}
\label{Tcons}
\ee
which vanishes due to the classical equations of motion and the
$X^+$ and $X^-$ constraints.  The key point
in (\ref{gzzEoM}) is that since $T_{01}$ and $\gamma _{01}$ vanish,
$T_{11}$ must be independent of sigma.  The auxiliary field equation
$T_{01}=T_{++}-T_{--}=0$ gives the full set of Virasoro constraints
except for the mass-shell condition. Evidently the non-zero mode part of
the $T^{(\gamma )}_{00}$ constraint only requires that the trace of the
stress tensor vanish.  This is the reason the theory must be Weyl
invariant.  Otherwise, the constraint is violated,
and the light-cone gauge fails.  Another way to put this is that the
light-cone gauge constraints $T^{(\gamma )}_{ab}=0$ are incompatible
with worldsheet general covariance $0=\nabla ^a T^{(g)}_{ab}=
\half \d _b \tr T^{(g)} + \nabla ^a T^{(\gamma )}_{ab}$ unless
the trace vanishes.  This problem is even more fundamental than
the fact that we cannot solve the constraints for $X^-$ if the
trace in non-zero.

At the classical level the trace of the stress tensor is
$\ap \Delta ^{(\gamma )} \Phi + e^{\phi} T$ (\ref{gammaEoM}).  This gets
renormalized upon quantization.  The classical trace may be cancelled
by the contribution from the Weyl anomaly.  It is the trace of the
renormalized stress tensor that must vanish in the quantum theory.
We will see how this works in section 3.

The zero mode part of the constraint gives the mass-shell condition.
The equation $T^{(\gamma )}_{00}=0$ may be solved for
$\dot{x}^- \equiv \d _{\tau} x^-$ in terms of the transverse $X^i$
fields.  The result is
\be
\dot{x}^- (\tau ) = - \frac{1}{2 \pi p^+} \int _0 ^{2\pi} d\sigma
\left\{ T^{tr \, (\gamma )}_{00} + p^+ G_{+i} \d _0 X^i - 2a \right\}
\label{mshell}
\ee
where the transverse stress tensor is
\be
T^{tr \, (\gamma )}_{00} = \half G_{ij} (\d _0 X^i \d _0 X^j
+ \d _1 X^i \d _1
X^j)+\half \ap ( -\frac{1}{p^{+\, 2}} \d _+^2 + \d _0 ^2  + \d _1 ^2)
\Phi .
\ee
We have introduced a
normal ordering constant, $a$, which vanishes classically, but is fixed
to a non-zero value in an anomaly free quantum theory.  Its
value is $(D-2)/24$, as determined by Lorentz invariance (if
present classically) or modular invariance.

It is interesting to see how the usual result for the $D=26$ critical
string arises.  It has the spacetime fields $B_{\mu \nu} = \Phi = T = 0$
and $G_{\mu \nu} = \eta _{\mu \nu}$.  The trace of the stress tensor
vanishes classically (\ref{gzzEoM}) and quantum mechanically (as is
well known in the conformal gauge, and will be checked within the
light-cone gauge in the next section).
Equation (\ref{mshell}) reduces to the usual mass-shell condition
\be
\dot{x}^- (\tau ) = - \frac{1}{p^+} (L_0^{tr} + \bar{L}_0^{tr} - 2).
\ee
where $L_0^{tr}$ ($\bar{L}_0^{tr}$) is the transverse part of the
zeroth left-moving (right-moving) Virasoro generator.  This gives
the mass $M^2 = p^+ p^- + p^i p^i = 2(N-1)$ where $N$
is the combined right and left oscillator number.  This is the correct
mass-shell relation for the critical string.  Equation (\ref{mshell})
generalizes the mass-shell for arbitrary target spaces.

The final constraint comes from varying the gauge fixed $X^+$.  This
variation is one of the terms in the stress tensor conservation equation
(\ref{Tcons}), which provides a convenient expression in terms of the
other fields
\be
\barcl
{\dst 0 = \frac{\delta S}{\delta X^+} } & = & {\dst
\frac{1}{2 \pi \ap p^+} \nabla ^a T_{a0}
- \frac{1}{p^+} \d _0 X^- \, \frac{\delta S}{\delta X^-}
- \frac{1}{p^+} \d _0 X^i \, \frac{\delta S}{\delta X^i}.}
\eac
\label{xplusEoM}
\ee
The first term is the divergence of the stress tensor which vanishes
if the theory is Weyl invariant in the light-cone gauge, as explained
above.  The second term is the $X^-$ constraint, and the third term
is proportional to the classical
equations of motion for the $X^i$ fields.  Thus, the $X^+$ constraint
is satisfied classically, given Weyl invariance.  This agrees with the
expectation that exactly four fields, $X^+, X^-, \gamma ^{00}$ and
$\gamma ^{01}$, should be eliminated in light-cone gauge fixing.  The
constraint $\frac{\delta S}{\delta X^+}$ is redundant.  It is
interesting to note that the $\gamma ^{00}$ constraint is almost trivial,
too.  Historically these constraints have not been emphasized
because they vanish classically when $\Phi = T = 0$, as in the
$D=26$ bosonic string.

Energy-momentum conservation carries over to the quantum theory.
So if the theory is Weyl invariant, the conservation equations
continue to hold.  In the conformal gauge it is possible to preserve
energy-momentum conservation even when the trace of the stress tensor
is non-zero.  The resulting Ward identities greatly ease the
calculation of the Weyl anomaly \cite{WItrick}.  This is not possible
in the light-cone gauge, since the gauge is only consistent when
the anomaly cancels.  There is no light-cone gauge for
non-critical strings, although it is possible to treat the
corresponding dilatonic critical strings.

\subsection{The Light-Cone Action}

This completes gauge fixing to the light-cone gauge classically.  It
remains to quantize the system and introduce string interactions.  The
gauge fixed form of the generating functional is
\be
\barcl
Z & = & {\dst \sum _{topologies} \int d\tau _r d\sigma _s
\int \DD X^i e^{iS[X^i]}}\\ [4mm]
{\dst S _{g.f.}} & = & {\dst \int d\tau \left\{ p^+ \dot{x}^-
+ \atepim \int _0 ^{2\pi}
d\sigma \left[ G_{ij} \d _a X^i \d ^a X^j - 2 p^+ G_{+i} \d _0 X^i -
(p^+)^2 G_{++} \right. \right. } \\
& & ~~~~~~~~~{\dst \left. \left.
+ B_{ij} \epsilon ^{ab} \d _a X^i \d _b X^j + 2 p^+
B_{+i} \d _1 X^i - \ap R^{(\gamma )} \Phi + T \right] \right\} .}
\eac
\label{lcaction}
\ee
$\tau _r$ and $\sigma _s$ are the worldsheet
moduli (the interaction times and locations) and
$R^{(\gamma )}$ has delta function support at the vertices.  We
have used Weyl invariance to set $\phi =0$.
The measure for the $X^i$ fields is flat, as we have checked at each
step of gauge fixing.

The Hamiltonian is also derived easily.  It is constructed as a Legendre
transform of the Lagrangian, as usual.  The canonical momentum densities
are given by
\be
P _i (\sigma ) = 2 \pi \ap \frac{\d L}{\d \dot{X}^i} =
G_{ij} \dot{X}^j + p^+ G_{+i} - B_{ij} \d _1 X^j .
\label{conjP}
\ee
Then the Hamiltonian is
\be
\barcl
H & = & {\dst \frac{1}{2 \pi \ap} \int _0 ^{2\pi} d\sigma
\left( \dot{X}^i P _i - \LL \right) } \\ [2mm]
& = & {\dst \atepi \int _0 ^{2\pi}
d\sigma \left[ G^{ij} P_i P_j + G_{ij} X^{i \, \prime} X^{j \, \prime}
- (p^+)^2 G^{--} + 2 p^+ P_i G^{i-} \right. } \\
& & {\dst ~~~~~ \left. + 2 p^+ B^-_{~i} X^{i \, \prime}
+ 2 G^{ij} P_j B_{ik} X^{k \, \prime} + G^{ij} B_{ik} B_{jl} X^{k \,
\prime} X^{l \, \prime} + \ap R^{(\gamma )} \Phi + T \right] }
\eac
\label{Ham}
\ee
where $\LL$ is the Lagrangian density (i.e. half the $\sigma$
integrand in (\ref{lcaction})).  The primes denote $\sigma$ derivatives.
Since $P^- (\sigma )$ is conjugate to $X^+ = p^+ \tau$, the Hamiltonian
is equal to $P^- (\sigma )/p^+$.

\section{Light-Cone Quantum Mechanics}
\setcounter{equation}{0}

The quantization of a general sigma model may be accomplished using
standard techniques extended to the light-cone gauge.  It is most
natural to use functional methods, since the light-cone operator
formalism is problematic.  There is a class of models which may
be solved exactly in the light-cone gauge, but most
actions with non-trivial space-time fields require the use of
perturbation theory on the worldsheet.  The resulting
string quantum mechanics is very complicated, due to the worldsheet
interactions.  That is, even without string interactions in
which the string branches and joins, the two-dimensional field theory
is interacting.  As with most interacting field theories, it is not
integrable, but at each order in perturbation theory we are able to
investigate whether the actions studied in section
\ref{subsec-GF} may be quantized consistently.

The usual technique for analyzing quantum sigma models is to
employ sigma model perturbation theory.  Then quantities such as the
effective action may be computed as power expansions in the coupling
constant $\sqrt{\ap} / r$, where $r$ is some relevant length scale (such
as the compactification radius).  We will not investigate whether
sigma model perturbation theory is compatible with the usual
light-cone techniques for calculating scattering amplitudes.  There
is no reason to expect the generalization of Neumann
functions and string field propagators not to make sense as series
expansions, but such considerations are beyond the scope of this
paper.  Many of the questions about the consistency of
quantization may be addressed within the context of string quantum
mechanics (without string interactions).  Sigma model perturbation
theory is perfectly natural in this milieu.

There are many approaches to sigma model perturbation theory that have
been developed in the conformal gauge.  Most of these also work in the
light-cone gauge with minor alterations.  A few techniques do not carry
over, so we will discuss the problems.  Our goal is to develop a framework
to test the consistency of the light-cone quantization,
not to explicitly calculate beta functionals or
effective actions using light-cone gauge. To that end, we will not be
concerned with selecting the best formulation for doing computations,
rather we will describe how to use the covariant techniques
within the light-cone gauge.  The essential point is that the Weyl
anomaly can be calculated, and it must vanish for the light-cone gauge
to be non-singular.

In this section we will quantize the general light-cone actions found in
the previous section.  The fields are quantized within sigma
model perturbation theory, either canonically or functionally.
The physical Hilbert space of states is
represented in terms of the transverse oscillators.  It consists of
all the mass eigenstates in the transverse Fock space.
Finally, the question of anomalies is
addressed.  The conformal anomaly manifests itself as an anomaly in the
Lorentz algebra when the target space has a Lorentz isometry.  In
general, the anomaly violates a gauge fixing constraint, so it must
vanish.

\subsection{Sigma Model Perturbation Theory}

The quantization of the gauge-fixed action using sigma model
perturbation theory begins with the division of the $X^{\mu}$ fields
into classical and quantum pieces, $X^{\mu}_0$ and
$\tilde{X}^{\mu}$, respectively.
\be
X^{\mu} (\sigma ^a ) = X^{\mu}_0(\sigma ^a ) + \tilde{X}^{\mu} (\sigma ^a).
\ee
Of course, in the light-cone gauge $X^+$ has no quantum part, and $X^-$
is largely irrelevant.
The counterterms necessary to renormalize the theory are covariant in
terms of the classical fields $X^i_0$, so it greatly simplifies
the analysis to write the action in an explicitly covariant form.
Covariant background field techniques are thoroughly explained in
the literature (in the conformal gauge) \cite{F}.
The main obstacle to a covariant action is that the naive quantum
fields are not covariant in spacetime. They need to
be expressed in terms of a spacetime vector. Then the fluctuations of
the metric and the other background fields are given by a
covariant expansion about their classical values.  The quantum fields
may be expressed covariantly in terms of geodesic coordinates
\be
\tilde{X}^{\mu} = x^{\mu} - \half \Gamma ^{\mu}_{~\sigma _1 \sigma _2}
x^{\sigma _1} x^{\sigma _2} + \cdots
\ee
where $x^{\mu}$ is tangent to the geodesic $\lambda ^{\mu}(t)$
running from $X^{\mu}_0$ at $t=0$
to $X^{\mu}_0 + \tilde{X}^{\mu}$ at $t=1$.  At arbitrary $t$, $\lambda
^{\mu}(t) = X^{\mu}_0 + x^{\mu} t - \half \Gamma ^{\mu}_{~\sigma _1
\sigma _2} x^{\sigma _1} x^{\sigma _2} t^2 + \cdots$, so $x^{\mu} = \d _t
\lambda ^{\mu}(0)$.  Note that the coordinates are defined separately at
each $\sigma ^a$, so $x^{\mu} = x^{\mu}(\sigma ^a)$.

Since the quantum field $x^{\mu}$ transforms covariantly, all of the
spacetime backgrounds are explicitly covariant in terms of
the classical field $ X^{\mu}_0$.  The metric is a particularly simple
example:
\be
G_{\mu \nu}(X) = G_{\mu \nu}(X_0)
- \frac{1}{3} R_{\mu \lambda \nu \rho}
(X_0) \, x^\lambda x^\rho - \frac{1}{6} \nabla_{\kappa}
R_{\mu \lambda \nu \rho}(X_0) \, x^{\kappa} x^\lambda x^\rho + \cdots
\label{bfe}
\ee
where $x^+=0$.
The light-cone gauge requirements are that $R_{-\lambda \nu \rho} =
B_{-\mu} = 0$ and $D_{-}$ annihilates any tensor.  These conditions are
true everywhere, including $X^{\mu}_0$, so $x^-$ does not appear.
The expansion of $\d _a X^{\mu}$ is
\be
\d _a X^{\mu} = \d _a X^{\mu} _0 + \nabla _a x^{\mu} + \frac{1}{3}
R^{\mu} _{~\lambda \kappa \nu } \d _a X^{\nu} _0 x^{\lambda} x^{\kappa}
+ \cdots
\ee
with $\nabla _a x^{\mu} = \d _a x^{\mu} + \Gamma ^{\mu} _{~\lambda
\kappa} (X_0) \d _a X^{\lambda}_0 x^{\kappa}$.  These expansions may be
derived using Riemann normal coordinates, but there are other tricks as
well (cf. \cite{Poly}).

The kinetic term for the $x^i$ fields turns out to be somewhat
complicated due to the classical backgrounds.  The spacetime metric
enters the quadratic term
\be
G_{\mu \nu}(X_0) \nabla _a x^{\mu} \nabla ^a x^{\nu}
\ee
which is messy because the metric $G_{\mu \nu}(X_0)$ is not constant.
The propagator
is simplified by introducing a vielbein $e^{\Is}_{\mu} (X_0)$ such that
\be
e^{\Is}_{\mu} (X_0) e^{\Js}_{\nu} (X_0) \eta _{\Is \Js}=G_{\mu \nu}(X_0).
\ee
Then $G_{\mu \nu}(X_0) \nabla _a x^{\mu} \nabla _b x^{\nu} = (\nabla _a
x) ^{\Is} (\nabla _b x) _{\Is}$ with $(\nabla _a x) ^{\Is}= \d _a x^{\Is}
+ \omega
^{\Is \Js}_{\mu}
\d _a X^{\mu}_0 x_{\Js}$ and $x^{\Is} = e^{\Is}_{\mu} (X_0) \, x^{\mu}$.
The spacetime spin connection one-form is given by
$\omega ^{\Is \Js}_{\mu} = e^{\Is}_{\mu} \nabla _{\lambda} e^{\mu J}$.
The kinetic term is $\d _a x^{\Is} \d ^a x^{\Is}$ up to spin
connection terms, so it gives a simple propagator.

We may now change
variables in the path integral
(\ref{SD}) from $X^{\mu}$ to $x^{\Is}$.  This involves a shift and a
spacetime coordinate change.  This change of variables is certainly
possible before light-cone gauge fixing,
since the measure may be assumed to be
invariant under spacetime diffeomorphisms and there is no Jacobian.
After the field redefinition gauge fixing may proceed as described in
section \ref{subsec-GF}.  Once the gauge has been fixed,
a field redefinition of $X^+$ could change the
form of the light-cone gauge conditions, so it is simplest to
make the field redefinition before fixing the gauge.

We are now ready to rewrite the action (\ref{lcaction}) in a background
field expansion.  The expansions for $G_{\mu \nu}, B_{\mu \nu}, \Phi, T$
and $\d _a X^{\mu}$ may be inserted in the action.
The $X^+$ field is gauge fixed, so it only appears in the classical
background field $X^{\mu}_0$.  $X^-$ has been eliminated as an
auxiliary field, as described in section \ref{subsec-GF}.
We will drop the tachyon and
anti-symmetric tensor backgrounds in order to simplify renormalization.
These backgrounds may be included using techniques well-developed in the
literature \cite{FT,CFMP}, but it simplifies our discussion to omit them.
Since $X^i_0$ satisfies the classical equations of motion, the
generating functional (without string interactions) becomes
\be
\barcl
Z_{X^i_0} & = & {\dst  \int \DD  x^{\Is} \, e^{-S[X^i(X^i_0, x^I)]}}\\
& = & {\dst \int \DD x^{\Is} \, e^{-S[X_0] -S_{X^i_0} [x^I]}}\\ [4mm]
{\dst S _{X^i_0}} & = & {\dst \int d\tau \, p^+ \dot{x}^-
+ \atepi \int d^2\sigma \left\{ (\nabla _a x)^{\Is} (\nabla ^a x)_{\Is}
+ R_{\mu \Is \nu \Js} (X_0) \, \d _a X^{\mu}_0 \d ^a X^{\nu}_0 x^{\Is}
x^{\Js}
\right. } \\ [2mm] & & ~~~~~~~~~{\dst \left.
+ \frac{4}{3} R_{\mu {\ssst I K J}} (X_0) \,
\d _a X^{\mu}_0 (\nabla ^a x)^{\Ks}
x^{\Is} x^{\Js} + \frac{1}{3} R_{{\ssst K I L J}} (X_0) \,
(\nabla _a x)^{\Ks}
(\nabla ^a x)^{\Ls} x^{\Is} x^{\Js}
\right. } \\ [2mm] & & ~~~~~~~~~{\dst \left.  - \ap
(-\d _a \d ^a \phi ) \left[ \Phi (X_0) + \nabla _{\Is} \Phi (X_0) \,
x^{\Is}
+ \nabla _{\Is} \nabla _{\Js}  \Phi (X_0) \, x^{\Is} x^{\Js} \right] +
\cdots \right\} }
\eac
\label{bflcact}
\ee
where the action has been Wick rotated ($\tau \ra i\tau$)
to a Euclidean worldsheet.  The ellipsis represents the terms
with derivatives of the Riemann tensor or higher derivatives of the
dilaton.  These higher order terms are given in reference \cite{CP},
along with an algorithm to generate them.
Any worldsheet curvature is due to the
the non-trivial Weyl mode.  The term linear in $x^{\Is}$ in the metric
expansion vanishes since $X_0$ is chosen to satisfy the
classical equations of motion with $\Phi =0$.  Since the dilaton breaks
Weyl invariance classically, it is put in the quantum part of the
action.  Note that there are no $X^+$ fluctuations.

The action (\ref{bflcact}) is easily quantized using
path integral techniques.  Collecting the quadratic terms gives the
full kinetic term
\be
\atepi \left[ \d _a x^I \d ^a x_I + 2 \omega _{\mu} ^{IJ} \d _a
X^{\mu}_0 x^J \d ^a x^I + \omega _{\mu} ^{IJ} \omega _{\nu} ^{IK}
\d _a X^{\mu} \d _a X^{\nu} x^J x^K + \ap \d _a \d ^a \phi
\nabla _I \nabla _J  \Phi (X_0) \, x^I x^J \right] ,
\ee
Actually, $\atepi \d _a x^I \d ^a x_I$ is used as the kinetic term,
with the
other terms considered as `mass' terms.  The spin connection is not
covariant, so it does not enter unless it is differentiated,
and the Liouville mode $\phi$ may be taken to be small and smooth.
Both types of terms may be treated perturbatively.  The propagator is
\be
\l x^{\Is} (\sigma , \tau ) x^{\Js} (\sigma \p , \tau \p) \r =
\ap \delta ^{{\ssst IJ}} \sum _{\stackrel {k^1= -\infty}{k^1 \ne 0}}
^{\infty} \int dk^0
\frac{e^{ik_a (\sigma ^a - \sigma ^{a\, \prime})}}{|k|^2}
\label{prop}
\ee
which is the kernel for the kinetic term on the cylinder.  At short
distances the propagator goes like
\be
\l x^{\Is} (\sigma , \tau ) x^{\Js} (\sigma \p , \tau \p) \r \sim
- 2\ap \delta ^{{\ssst IJ}}
\log \left| \rho - \rho \p \right| ~~~~~{\rm as~}\rho \ra \rho \p
\label{shortprop}
\ee
This is the usual free field propagator for a flat worldsheet.  The
worldsheet curvature is treated perturbatively.

\subsection{The Hilbert Space of States}

The fields have mode expansions given by
\be
x^I (\sigma , \tau ) = 2 \ap p^I \tau
+ i\sqrt{2\ap} \left( \sum _{n=-\infty} ^{\infty} \alpha ^I _n e^{n\rho}
+ \sum _{n=-\infty} ^{\infty} \alphabar ^I _n e^{n\rhobar} \right)
\label{modeexp}
\ee
where $\rho = \tau + i\sigma$.  The oscillators satisfy the
commutation relations
\be
\barcl
{\dst [\alpha ^I_m, \alpha ^J _n ]} & = &
{\dst m \, \delta ^{{\ssst IJ}} \delta _{m, -n}} \\
{\dst [\alphabar ^I_m, \alphabar ^J _n ]} & = &
{\dst m \, \delta ^{{\ssst IJ}} \delta _{m, -n}} \\
{\dst [\alpha ^I_m, \alphabar ^J _n ]} & = & 0
\eac
\label{commrel}
\ee
All of the gauge symmetry has been fixed by the light-cone gauge, so the
physical Hilbert space of states is represented by the whole Fock space
of transverse oscillators acting on the vacuum labeled by $p^\mu$
\be
\barcl
\Hphys & = & {\dst span \left\{ \prod _{I=1}^{D-2}
\left( (\alpha ^I_{-N})^{m_{I,N}}
\cdots (\alpha ^I_{-1})^{m_{I,1}} \right) \left( (\alphabar
^I_{-N})^{\mbar _{I,N}}\cdots (\alphabar ^I_{-1})^{\mbar _{I,1}} \right)
| p^\mu \r + \OO (\ap ) \right. } \\
& & {\dst \left. {\rm ~~~~~~~such~that~} \sum _{I,n} n\, m_{I,n} =
\sum _{I,n} n\, \mbar _{I,n} = N {\rm ~for~} N=0,1,2,\cdots
\right\} }
\eac
\label{Hilbert}
\ee
such that $p^\mu$ satisfies the mass-shell condition (\ref{mshell}).
The physical states must be $L_0^{tr}$ and $\Lbar _0^{tr}$
eigenstates for the mass-shell equation to have a solution.  The
monomials in the oscillators shown in equation (\ref{Hilbert}) are
eigenstates at $\ap =0$, but must be corrected for $\ap \ne 0$.  These
corrections may change the mass spectrum (the eigenvalues) as well.
Note that the left- and right-moving oscillator numbers are equal, as
required by $\sigma$ translation invariance for the closed string.
The vacuum, $| p^\mu \r$, is annihilated by the positive frequency
modes of the string,
$\alpha ^I_{n}| p^\mu \r = \alphabar ^I_{n}| p^\mu \r =0$ for $n>0$, for
all $\ap$.
The Hilbert space is also endowed with an algebraic structure given by
the three-point functions of the states.  Even though the cardinality of
the Hilbert space is determined by the number of transverse dimensions,
not all actions with the same number of dimensions have isomorphic
Hilbert spaces.  Both the mass spectrum and the three-point functions
may be computed in sigma model perturbation theory.

\subsection{Renormalization and the Beta Functions}

The action (\ref{bflcact}) suffers from
both quadratic and logarithmic divergences in the UV.  The propagator
diverges logarithmically at short distances, so any contraction of two
of the quantum fields in the action is potentially problematic.
The quadratic divergences
reflect the presence of the tachyon.  They renormalize the cosmological
constant of the two dimensional quantum gravity.  It is the logarithmic
divergences that can destabilize the light-cone gauge.
The action must be renormalized in a way consistent with worldsheet
reparameterization invariance, such as dimensional regularization with
minimal subtraction.  In the process the spacetime fields get
renormalized.  The dimension two metric operator undergoes an additive
renormalization, and the dimension zero dilaton receives both
multiplicative and additive renormalizations due to the worldsheet
curvature.  Even though the action may be Weyl invariant classically,
the renormalized
worldsheet couplings (i.e. the physical spacetime metric, antisymmetric
tensor field, etc.) may vary with the Weyl scale, since there is no
regulator that is both reparameterization and Weyl invariant.
Unless the Weyl anomaly vanishes, the light-cone gauge is not
consistent.

The light-cone quantum mechanics that is emerging is remarkably similar
to the sigma model perturbation theory in the conformal gauge.
The action is identical,
except for the absence of the kinetic term $\d _a x^+
(\nabla ^a x)^-$ and the interaction terms with $x^+$.  Due to the
light-cone restrictions on the backgrounds, $x^-$ would not appear in
the interaction terms.  Since $x^+$ would only contract with $x^-$,
the sole divergent term missing from the light-cone action is the
kinetic term itself (which contributes to the dilaton beta function).
Except for this possible difference in $\beta ^{\Phi}$ at leading order,
the beta functions must be identical.  The only caveat is
that not all of the calculational methods used in the conformal gauge
are appropriate for the light-cone gauge.

The standard renormalization of the divergent sigma model actions uses
dimensional regularization with minimal subtraction.  This is how we
will renormalize the light-cone action (\ref{bflcact}).  In
$2+\epsilon $ dimensions the action is given by
\be
\barcl
{\dst S _{X^i_0}} & = & {\dst \atepi \int d^{2+\epsilon}\sigma
\, e^{\half \epsilon \phi} \, 2 p^+ \d _0 X^-[X^{\mu}_0,x^I]
} \\ [2mm] & & ~~{\dst
+ \atepi \int d^{2+\epsilon}\sigma \left\{
e^{\half \epsilon \phi} \left[ (\nabla _a x)^I (\nabla ^a x)_I
+ R_{\mu I \nu J} (X_0) \, \d _a X^{\mu}_0 \d ^a X^{\nu}_0 x^I x^J
\right. \right. } \\ [2mm] & & ~~~~{\dst \left.
+ \frac{4}{3} R_{\mu I K J} (X_0) \, \d _a X^{\mu}_0 (\nabla ^a x)^K
x^I x^J + \frac{1}{3} R_{K I L J} (X_0) \, (\nabla _a x)^K
(\nabla ^a x)^L x^I x^J \right]
} \\ [2mm] & & ~~~~{\dst \left.  - \ap
(-\d _a \d ^a \phi - \frac{\epsilon}{4} \d _a \phi \d ^a \phi)
\left[ \Phi (X_0) + \nabla _I \Phi (X_0) \, x^I
+ \nabla _I \nabla _J  \Phi (X_0) \, x^I x^J \right] + \cdots \right\} }
\eac
\label{dract}
\ee
where the worldsheet curvature to which the dilaton couples is taken to
be the dimensional continuation of $\sqrt{g} R^{(2)}/(d-1)$, for
convenience.  It is now clear that the regulated form
of the action is not Weyl invariant.  The Weyl scale ($g_{ab} = e^\phi
\delta _{ab}$) enters through the
worldsheet volume element $\sqrt{g}$ as well as the curvature.

The action (\ref{dract}) has been split into light-cone and transverse
parts, because the transverse part is almost identical to the
corresponding part of the action in the conformal gauge.  The light-cone
part is more unusual.  Because of the Weyl mode $\phi$, the $X^-$
oscillators reenter the action in $2+\epsilon $ dimensions
\be
\int d\tau \, 2 p^+ \dot{x}^- \ra
\int d\tau \, 2 p^+ \dot{x}^- +
\atepi \int d^{2+\epsilon}\sigma \, \epsilon \phi  p^+
\d _0 X^- +\cdots .
\ee
The classical solution for $X^-$ in terms of the transverse fields is
singular, with poles as $\epsilon \ra 0$, so this term may contribute to
the Weyl dependence of the renormalized action.

The form of the trace of the stress tensor is given in (\ref{Ttrace})
for the transverse action which is covariant on the worldsheet
\be
\sqrt{g} \, T^a _{\, a} =
\beta ^T (X) \, \sqrt{g} +
\beta ^{\Phi} (X) \, \sqrt{g} \, R^{(2)} +
\beta ^G_{\mu \nu} (X) \, \sqrt{g} \d _a X^{\mu} \d ^a X^{\nu} +
\beta ^B_{\mu \nu} (X) \, \epsilon ^{ab} \d _a X^{\mu} \d _b X^{\nu}.
\label{Ttrtwo}
\ee
This may be integrated to get the $\phi$-dependent part of the effective
action,
\be
\barcl
{\dst S_{eff}[\phi] } & = & {\dst
 \int d^2 \sigma \left\{ \beta ^T (X)
\, e^\phi + \half \beta ^{\Phi} (X) \, \d _a \d ^a \phi ^2
\right.} \\
& & {\dst ~~~~~ \left.
+ \beta ^G_{\mu \nu} (X) \, \phi \delta ^{ab} \d _a X^{\mu} \d _b X^{\nu}
+ \beta ^B_{\mu \nu} (X) \, \phi \epsilon ^{ab} \d _a X^{\mu} \d _b
X^{\nu} \right\} , }
\eac
\ee
where we have dropped a term proportional to the scalar curvature that
is related to $\beta ^{\Phi}$.
The corresponding terms are easily extracted from the action
(\ref{dract}).  The exponential $e^{\half \epsilon \phi}$ is expanded,
and a field redefinition $x^I = (1 - \frac{1}{4} \epsilon \phi ) y^I$
is performed to simplify the propagator \cite{GSW}.  This is related to a
required wave function normalization.  The poles in $\epsilon$ in the
action are discarded through minimal subtraction, leaving
finite terms which contribute to the Weyl anomaly.
The Weyl anomaly coefficients are read off using
\be
\l x^\Is x^\Js \r \sim - \frac{\ap}{2 \epsilon} \delta ^{{\ssst IJ}},
\ee
which is the singular part of the propagator.

Consider first this transverse part of the action.  It is identical to
the corresponding part of the action in conformal gauge, as explained
above, so using the methods we have just outlined the anomaly
coefficients may be computed identically in both cases.  The only
difference is the leading order part of the dilaton beta function, which
counts the number of dimensions minus the ghost contribution.  In the
conformal gauge it is proportional to $(26-D)$, whereas in the
light-cone gauge it is $(2-D)$ due to the absence of the propagating
ghosts and the light-cone coordinates.  The resulting beta functions
are
\be
\barcl
(\beta^G_{\mu\nu})^{tr}
&=&{\dst R_{\mu\nu} +2\nabla _{\mu}\nabla _{\nu}\Phi
+ \cdots }\\
(\beta^{\Phi})^{tr}
&=&{\dst -\frac{\ap}{16\pi ^2}\left[ \frac{2-D}{3\ap} +
 R-4\d _{\mu}\Phi\d ^{\mu}\Phi+4\nabla ^2\Phi +\cdots \right]}.
\eac
\label{lctrbfns}
\ee
At this point we have neglected the two contributions to the Weyl anomaly
that appear in the light-cone gauge differently from the conformal
gauge.  The first contribution comes from the dimensional continuation
of the $\int 2 p^+ \dot{x}^-$ term, and the second comes from the
measure.  Together they eliminate the discrepancy in the dilaton beta
function.

First consider the $X^-$ term.  Expanded out in terms of the transverse
fields, it is given by
\be
\bal
{\dst \atepim \int d^{2+\epsilon}\sigma \left\{
\, (e^{\half \epsilon \phi}-1) \, \int ^\sigma d\sigma \p \, \d _0
\left[ (\nabla _0 x)^I (\nabla _1 x)_I
+ R_{\mu I \nu J} (X_0) \, \d _0 X^{\mu}_0 \d _1 X^{\nu}_0 x^I x^J
\right. \right. } \\ [2mm] ~~~~~~~~~{\dst \left.
+ \frac{2}{3} R_{\mu I K J} (X_0) \, [\d _0 X^{\mu}_0 (\nabla _1 x)^K
+ \d _1 X^{\mu}_0 (\nabla _0 x)^K ] x^I x^J
+ \frac{1}{3} R_{K I L J} (X_0) \, (\nabla _0 x)^K
(\nabla _1 x)^L x^I x^J \right]
} \\ [2mm]  ~~~~~~~~~{\dst \left.  + \ap
(\nabla _0 \nabla _0 + \d _0 \d _0 \phi )
\left[ \Phi (X_0) + \nabla _I \Phi (X_0) \, x^I
+ \nabla _I \nabla _J  \Phi (X_0) \, x^I x^J \right] + \cdots \right\} .}
\eac
\ee
The terms contributing to the Weyl anomaly may be organized in much the
same way as with the transverse part of the action.  After all, $X^-$
is proportional to $T^{tr}_{01}$, which is the variation of the action
with respect to $\gamma ^{01}$.  The Weyl anomaly takes the form
\be
\barcl
{\dst (S_{eff}[\phi])^{(X^-)} } & = & {\dst
\int d^2 \sigma \left\{
\half \left[ \beta ^{\Phi} (X)\right] ^{tr} \, \d _0 \d _0 \phi ^2
+ \phi \d _0 \int ^{\sigma} \left[\beta ^G_{\mu \nu} (X)\right]^{tr}
\, \d _0 X^{\mu} \d _1 X^{\nu}
\right\} , }
\eac
\ee
up to terms that vanish because $X^{\mu}_0$ satisfies the classical
equations of motion.  The transverse part is actually the whole metric
beta functional, so it will be required to vanish.  The dilaton beta
functional is incomplete, and the $X^-$ contribution merely cancels the
part of the Weyl anomaly that depends on $\tau$ derivatives of $\phi$.

The remaining part of the anomaly is cancelled by the measure.  There
are two contributions to the measure:  one from the Faddeev-Popov
determinant (\ref{FPdet}) and one from the Jacobians and from solving
for $X^-$ (\ref{Xminus}) and $\gamma ^{01}$ (\ref{gzero}).  Dropping
the factors of $p^+$ which cancel, the two contributions are
\be
\asize{1.0}
\barcl
\Delta _{FP} & = & {\dst \left|
\bacc
1 & 0 \\
\d _0 & - \d _1
\eac \right| = \left[ \det (\d _1^2)_{vector} \right] ^{1/2} } \\ [2mm]
\JJ & = & {\dst \det (\d _1 )
= \left[ \det (\d _1^2)_{scalar} \right] ^{-1/2} }
\eac
\asize{1.4}
\ee
where the subscripts {\em vector} and {\em scalar} refer to the fields
on which the operators act.  Specifically, $(\d _1^2)_{vector}$ acts on
the $\sigma$ component of the reparameterization vector.  The
expressions involving $\det (\d _1^2)$ are derived using the covariant
measure on the space of metrics (\ref{metmeas}) and the measure on the
space of fields $X^i$. These
determinants may be evaluated in sigma model perturbation theory, giving
the contribution
\be
\left( \Delta _{FP} \, \JJ \right) _{\phi} =
-\frac{1}{2\pi ^2} \int d^2 \sigma \d _1 \d _1 \phi ^2 + \mu e^{\phi}
\ee
This combines with the contributions from the transverse action and the
$X^-$ term to produce the complete beta functionals
\be
\barcl
\beta^G_{\mu\nu}
&=&{\dst R_{\mu\nu} +2\nabla _{\mu}\nabla _{\nu}\Phi
+ \cdots }\\
(\beta^{\Phi})^{tr}
&=&{\dst -\frac{\ap}{16\pi ^2}\left[ \frac{26-D}{3\ap} +
 R-4\d _{\mu}\Phi\d ^{\mu}\Phi+4\nabla ^2\Phi +\cdots \right]}.
\eac
\label{lcbfns}
\ee
These are the usual beta functionals.
They correspond to the light-cone form of the Weyl anomaly
\be
\barcl
{\dst S_{eff}[\phi] } & = & {\dst \frac{1}{4\pi}
 \int d^2 \sigma \left\{ \beta ^T (X)
\, e^\phi + \beta ^{\Phi} (X) \, \phi \d _1 ^2 \phi
+ f (X) \d _1 \d _1 \phi \right.} \\
& & {\dst ~~~~~ \left.
+ \beta ^G_{\mu \nu} (X) \, \phi \delta ^{ab} \d _a X^{\mu} \d _b X^{\nu}
+ \beta ^B_{\mu \nu} (X) \, \phi \epsilon ^{ab} \d _a X^{\mu} \d _b
X^{\nu} \right\} , }
\eac
\label{effSphi}
\ee
where we have reinstated the tachyon and anti-symmetric tensor for
completeness.  This differs from the usual effective action for the
Liouville mode.  $\phi$ does not propagate, so (\ref{effSphi})
cannot be quantized in any naive fashion to yield a non-critical
light-cone string.

This calculation of the Weyl anomaly has identified the
critical dimension for light-cone string theory without resorting to
computing the Lorentz anomaly.  The Weyl anomaly is the more fundamental
of the two, since it can exist even when the sigma model does not have a
global Lorentz isometry.  The light-cone gauge cannot be fixed if the
anomaly does not vanish, as we found in section \ref{subsec-GF}.  One
might ask what happens if the theory is formulated violating this
constraint.  After all, the light-cone action (\ref{lcaction}) still
exists.  The first problem is that the Weyl mode does not decouple, so
it must be quantized, too.  If one sets $\phi =0$ by fiat, the action
is not renormalizable.  Also, the underlying gauge invariance that
insures proper factorization of loop graphs would be spoiled, and the
theory would be inconsistent once interactions were included.  The
measure at the vertices would be wrong.  The Weyl
anomaly must vanish for the light-cone gauge to be consistent.

Before we proceed to introduce interactions, we will point out one
technique for computing the Weyl anomaly that does not work in
light-cone gauge.  It is the trick of using conformal Ward identities to
compute the dilaton beta functional on a flat worldsheet \cite{WItrick}.
Since the
$\beta ^{\Phi}$ term in the effective action is quadratic in $\phi$,
the classical two-point function
\be
\l \frac{\delta ~}{\delta \phi (\rho )} \frac{\delta ~}{\delta \phi
(\rho \p )} \r
= \l T^a _{\, a} (\rho)  T^a _{\, a} (\rho \p) \r ,
\ee
contains the $\beta ^{\Phi}$ information.  In conformal field theory,
this is related to the two-point function of the holomorphic part of the
stress tensor via conformal Ward identities, $\l \nabla ^a T_{ab} \cdots
\r =0$, allowing an extremely simple computation of the leading term of
the dilaton beta functional in terms of the central charge.
Unfortunately, these Ward
identities are violated in the non-critical light-cone gauge systems.
The  holomorphic part of the stress tensor is being set to
zero when we solve for $X^-$, but the trace does not vanish.
The Ward identities do not hold, and without the use of the Ward
identities, we must resort to the calculation of the critical dimension
using a curved worldsheet, as we have done above.  Once the leading term
is calculated, the higher order terms may be found using the elegant
method of Curci and Paffuti \cite{CP}.  It uses beta function
consistency conditions to compute the dilaton beta function on a flat
worldsheet.

\section{Interactions and Exactly Solvable Models}
\label{subsec-Int}
\setcounter{equation}{0}

Once we have a light-cone gauge theory that is consistent at the level
of quantum mechanics, the next step is to see whether we can add string
interactions.  An interacting string theory allows the calculation of
N-string scattering amplitudes in a string loop perturbation theory.  In
the first quantized formalism, the amplitude at a fixed order in the
string coupling is expressed as an integral
over the moduli of a Riemann surface representing the geometry of the
N-string scattering process.  Each Riemann surface has a definite
geometry in which strings split and join at vertices.  Since we are
considering closed string theory, the basic string vertex is one in
which two strings join to form a single string, or the time-reversed
vertex in which one string splits to form two.  This is the type of
string interaction that must be added (possibly with contact terms as
well).

\subsection{String Interactions}

Since the worldsheet is no longer just a cylinder, we must reconsider
one aspect of gauge fixing.  The equations of motion for $X^+$ may not
be satisfied by the previous gauge choice.
For simplicity we will consider tachyon
scattering.  The backgrounds are taken to satisfy the Weyl invariance
conditions and the light-cone gauge requirements discussed above.
The N-tachyon scattering amplitude
before gauge fixing is given by
\be
\renewcommand{\arraystretch}{1.7}
\barcl
\l  X^{\mu}_1 \cdots  X^{\mu}_N \r & = & {\dst \sum _{loops} \int d\mu
\int [\DD g_{ab}] \, \DD X^{\mu} e^{-S[g_{ab},X]}
\delta (X^{\mu}(\sigma ^a_1) -X^{\mu}_1) \cdots
\delta (X^{\mu}(\sigma ^a_N) -X^{\mu}_N)}\\
\l  p^{\mu}_1 \cdots  p^{\mu}_N \r
& = & {\dst \sum _{loops} \int d\mu
\int [\DD g_{ab}] \, \DD X^{\mu} e^{-S[g_{ab},X]}
e^{ip^{\mu}_1 X_{\mu}(\sigma ^a_1)}  \cdots
e^{ip^{\mu}_N X_{\mu}(\sigma ^a_N)} }\\
S & = & {\dst \atepi \int d^2\sigma \, \sqrt{g}
\left\{ g ^{ab} G_{\mu \nu}\d _a X^{\mu} \d _b X^{\nu}
+\frac{\epsilon ^{ab}}{\sqrt{g}} B_{\mu \nu}\d _a X^{\mu} \d _b X^{\nu}
- \ap R^{(2)} \Phi + T
\right\} }
\label{Nstramp}
\eac
\renewcommand{\arraystretch}{1.4}
\ee
where the sum over loops is a sum over the genera of the Euclidean
worldsheet.  The integral over $\mu$ sums over the moduli space of
N-punctured surfaces at fixed genus, including the integrals
$\int d^2 \sigma ^a_r \sqrt{g}$ over the location of the punctures.
We will describe the modular integral
more precisely below.  The N-tachyon momentum amplitude is just the
Fourier transform of the position amplitude.  The exponentials must be
normal ordered in the quantum theory \cite{PolNO}.  Scattering of the
higher string states may be treated similarly using well-known tricks.

Classically, we may pull the $X^+$ and $X^-$ parts of the exponentials
into the action, giving
\be
\barcl
S \p & = & {\dst \atepi \int d^2\sigma \, \sqrt{g}
\left\{ g ^{ab} G_{\mu \nu}\d _a X^{\mu} \d _b X^{\nu}
+\frac{\epsilon ^{ab}}{\sqrt{g}} B_{\mu \nu}\d _a X^{\mu} \d _b X^{\nu}
- \ap R^{(2)} \Phi
\right. } \\ & & ~~~~~~~~~~~~~~~~~~~~~~~~~ {\dst \left.
+ T + i \sum _{r=1}^N ( p^+ _r X^- + p^- _r X^+ )
\right\} .}
\eac
\ee
Since all the background fields are independent of $X^-$, the equation
of motion for $X^+$ is
\be
\Delta X^+ = i \sum _{r=1}^N p^+ _r (1/\sqrt{g}) \,
\delta ^{(2)} (\sigma ^a - \sigma
^a_r) .
\label{NptEoM}
\ee
This has the simple solution $X^+ = \tau$, having let $\tau _r \ra \pm
\infty$.  The factor of $p^+$ in the previous gauge condition is
dropped since different string legs have different values of $p^+$, and
$X^+$ must be consistent around loops.  Each leg is rescaled compared to
the cylindrical worldsheet by its value of
$p^+$ ($\sigma \ra \sigma/p^+ , \tau \ra \tau /p^+$).  Since $\sigma$
runs from $-2\pi |p^+|$ to $2\pi |p^+|$,
the canonical momentum is still $p^+$.
Figure 1 shows the
worldsheet after the rest of the gauge fixing has been carried out.
This is called a Mandelstam diagram.  Note that the width is constant
due to $p^+$ momentum conservation.
The essential property of the Mandelstam diagram is that it puts the
$X^-$ vertex operators in the infinite past or the infinite future, so
that the ill-defined object $e^{ip^+_r X^-}$ does not contribute except
through the width of the external legs.

\begin{figure}[tb]
\centering
\parbox{4.6in}{
{\epsfxsize=4.5in\epsfbox{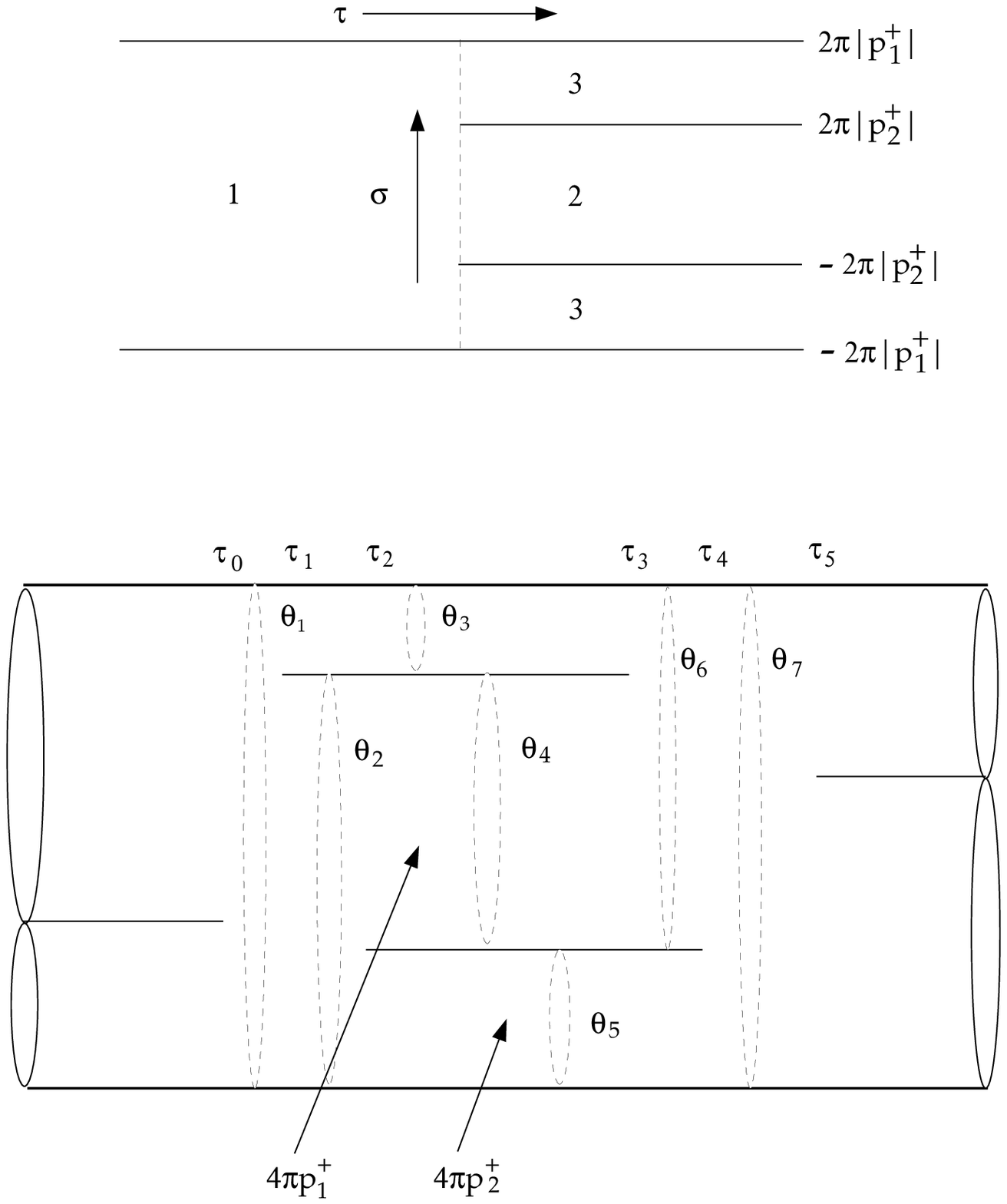}}
}
\renewcommand{\baselinestretch}{1.0}
\parbox{5.2in}{
{\small Figure 1:~~The three string vertex and a two-loop,
four-string Mandelstam diagram with the moduli shown.}}
\renewcommand{\baselinestretch}{1.3}
\end{figure}

The geometry of the Mandelstam diagram is that of cylinders joined by
three string vertices.  There is a delta function singularity in the
curvature at the vertex which contributes $-1$ to the Euler character of
the worldsheet.  This curvature would enter the $X^+$ equation of motion
if the dilaton depended on $X^-$. Then $X^+ = \tau$ would no longer be a
solution.  For this reason, as well as to prevent ghosts, we have
required $\d _- \Phi =0$.
This point must be considered further in the treatment of $c=1$
\cite{c1gr}.

The moduli of the Mandelstam diagram are shown in Figure 1.
For an $N$-string scattering diagram with $g$ loops
(genus $g$), they consist of the $g$ internal $p^+$ momenta which
determine the radii of the internal legs, the $N-2+2g$ interaction times
and the $N-2+3g$ angles, modulo the global time-translation and
$\sigma$-rotation invariance.  Thus, there are $2(N-3+3g)$ real moduli,
as required.

The connection between the Mandelstam diagrams and the non-singular
parameterizations of Riemann surfaces may seem obscure at this point.
 From the point of view of physics, the connection is very simple,
especially at tree level.  Consider gauge fixing on the complex plane
with coordinates $z$ and $\zbar$, where the
interaction points $z_r$ are at a finite distance from the origin.
The solution to the classical
equation of motion (\ref{NptEoM}) is given by
\be
X^+ =  \sum _{r=1}^N 2 p^+ _r \log | z - z_r |
\ee
Since $X^+$ satisfies Laplaces equation everywhere except the points $z_r$,
we may consider the conformal transformation, $z \ra \rho$, where
\be
\rho (z) = \sum _{r=1}^N 2 p^+ _r \log ( z - z_r ).
\label{SC}
\ee
The dilaton and tachyon are independent of $X^-$, so $X^+$
transforms essentially as a weight zero field under conformal
transformations, and $X^+ = \RR e (\rho)$ in the new coordinates.
We may set $\rho =
\tau + i\sigma $, so that the light-cone gauge condition is
reproduced.  The transformation (\ref{SC}) is called a
Schwarz-Christoffel transformation, and it maps the N-punctured plane
onto the Mandelstam diagram exactly once.  The turning points of the
Schwarz-Christoffel map occur at the vertices of the Mandelstam surface,
so the solutions of $\d _z \rho =0$ relate the light-cone moduli to the
Koba-Nielson variables $z_r$.

Thus, we have shown that the light-cone gauge is well-defined on
Mandelstam diagrams provided the backgrounds meet the requirements for a
consistent quantum mechanics.  We could proceed to formulate scattering
amplitudes for these backgrounds using a double perturbation series in
both $\ap$ and the string coupling, but instead we consider backgrounds
which are exactly solvable at the worldsheet level and do not have
contributions at higher order in $\ap$.

\subsection{Exactly Solvable Models}

Consider the light-cone action (\ref{lcaction})
\be
\barcl
{\dst S _{g.f.}} & = & {\dst \int d\tau \left\{ p^+ \dot{x}^-
+ \atepi \int
d\sigma \left[ G_{ij} \d _a X^i \d ^a X^j - 2 p^+ G_{+i} \d _0 X^i -
(p^+)^2 G_{++} \right. \right. } \\
& & ~~~~~~~~~{\dst \left. \left.
+ B_{ij} \, \epsilon ^{ab} \d _a X^i \d _b X^j + 2 p^+
B_{+i} \d _1 X^i - \ap R^{(\gamma )} \Phi + T \right] \right\} .}
\eac
\ee
This action is exactly solvable in terms of the classical equations of
motion provided it is at most quadratic in the fields
$X^i$.  The backgrounds must satisfy the requirements
\be
\bac
\d _k G_{ij} = \d _k B_{ij} = 0 \\
\d _k \d _j G_{+i} = \d _k \d _j B_{+i} = 0 \\
\d _k \d _j \d _i G_{++} = \d _k \d _j \d _i T = 0
\eac
\label{intconds}
\ee
since this guarantees that the transverse coordinates enter at most
quadratically.  The dilaton only contributes at the vertices (where the
curvature is located), so it does not affect the solvability.

Since the $X^+$ dependence of the spacetime fields is not constrained by
solvability, there is potentially a large class of backgrounds for
which the sigma models are exactly solvable in the light-cone gauge.
Many of these sigma models would be non-trivial in
the conformal gauge.  There are certainly some configurations of the
metric and dilaton for which this is true, since the dilatonic
gravitational waves studied in ``Compactification Propagation''
\cite{comprop} fall into this class.

In fact, the techniques of that paper \cite{comprop} are easily extended
to find all solutions of the background equations with $T=0$ that meet
the solvability conditions.  The metric beta function with $T=0$ forces
$\Phi$ to be independent of the transverse coordinates, up to a linear
term, $Q^iX^i$.  All the
beta functions except $\beta ^G_{++}$ vanish automatically in the
critical dimension
dimension.  $\beta ^G_{++}=0$ may be solved for $\Phi (X^+)$:
\be
\Phi = \frac{1}{8} \int ^{X^+} \int \left\{
G^{ik} G^{jl} H_{+ij} H_{+kl} - 4 R_{++} \right\} + Q^iX^i
\ee
because the integrand only depends on $X^+$.  Since the $X^+$ dependence
of the metric and the antisymmetric tensor field is unconstrained, there
is indeed a large class of solutions.  Each of these yields an exactly
solvable  sigma model in the light-cone gauge.

The spacetime fields only depend on $\sigma$ through the
coordinates $X^i$, giving a diagonal
action in the oscillator basis.  It
simplifies the field theory to consider fields that are asymptotically
constant functions of $X^+$, so that the potentials are flat at
infinity.  The
worldsheet has been rescaled, so the oscillator expansion of $X^i$ in
(\ref{modeexp}) becomes
\be
X^i (\sigma , \tau ) = x^i (\tau )
+ i\sqrt{2\ap} \left( \sum _{n=-\infty} ^{\infty} \alpha ^i _n (\tau )
e^{in\sigma /(2p^+)}
+ \sum _{n=-\infty} ^{\infty} \alphabar ^i _n (\tau )
e^{in\sigma /(2p^+)} \right)
\label{oscexp}
\ee
The spacetime fields may be decomposed explicitly in terms of the
transverse coordinates
\be
\barcl
G_{+i} & = & {\dst G_{+i}^{(0)} (\tau) + [G_{+i}^{(1)}(\tau)]_j \, X^j } \\
B_{+i} & = & {\dst B_{+i}^{(0)} (\tau) + [B_{+i}^{(1)}(\tau)]_j \, X^j } \\
G_{++} & = & {\dst G_{++}^{(0)} (\tau) + [G_{++}^{(1)}(\tau)]_j \, X^j
+ [G_{++}^{(2)}(\tau)]_{jk} \, X^j X^k} \\
T & = & {\dst T^{(0)} (\tau) + [T^{(1)}(\tau)]_j \, X^j
+ [T^{(2)}(\tau)]_{jk} \, X^j X^k}
\eac
\ee
where we have set $X^+ = \tau$.  Because the spacetime fields only
depend on $\sigma$ through the $X^i$ fields, the terms in the action
linear in the transverse fields drop out.  The resulting form of the
action is
\be
S _{int} = \frac{p^+}{2\ap} \int d \tau \,
A^{i\, T}_n \cdot M_{ij} (\tau ) \cdot A^{j}_{-n}
\ee
with
\be
\asize{1.0}
A^i _n (\tau ) = \left( \bac \dot{\alpha} ^i_n \\ \dot{\alphabar}^i_n \\
\alpha ^i_n \\ \alphabar ^i_n \eac \right) ~~~ {\rm and~~~}
M_{ij} (\tau ) = G_{ij} (\tau ) \left( \bacc -1 & 0 \\ 0 & n^2 \eac
\right) + \cdots .
\asize{1.4}
\ee
This is a slightly unusual form for a quadratic action, because the
kinetic term is multiplied by a time-dependent function.  It is still
exactly solvable, however.  The propagator is entirely determined by the
classical equations of motion for the $X^i$ fields.
We will not display the propagator explicitly,
since it is a messy expression whose exact form does not affect what
follows.  Horowitz and Steif \cite{HorSt} have used an approximate form
of the propagator to examine the excitation of a string passing through
a plane-fronted wave at lowest order in the string coupling.  The
Bogoliubov transformation is easily calculated once the propagator is
known.

Once the string propagator is determined in the oscillator basis, all
that remains is to determine the vertex in that basis.  The only vertex
that is necessary for the $D=26$ closed bosonic string field theory is
the three string vertex.  No contact terms are necessary.  This
vertex is simply an overlap delta functional in the position
representation, giving the decomposition in the oscillator basis
\cite{Mand1,LCSFT}
\be
\barcl
|V\r & = & {\dst \exp \left\{
- \tau _0 \sum _{r=1}^3 1/(2p^+_r) + \half \sum
_{r,s} \sum _{m,n=1}^{\infty} \Nbar _{mn} ^{\, rs} \alpha _{-m}^r \cdot
\alpha _{-n}^s \right. } \\
& & ~~~~~~~~~ {\dst \left.
+ \sum _{r,s} \sum _{m=1}^{\infty} \Nbar _{mn} ^{\, rs}
\alpha _{-m}^r \cdot \PP  - \frac{\tau _0}{16 p^+_1 p^+_2 p^+_3} \PP ^2
\right\} | 0 \r \, \delta ^{(D-2)} (\sum _r p_r) }
\eac
\ee
where $\tau _0$ is the interaction time,
$|0\r$ is the oscillator vacuum, $\PP ^i = 2( p^+_1 p_2^i - p^+_2
p_1^i)$, $\Nbar _{mn} ^{rs} = N_{mn} ^{rs} e^{m\tau _0 /(2p^+_r)}
e^{n\tau _0 /(2p^+_s)}$ and $\Nbar ^r_m = N ^r_m  e^{m\tau _0
/(2p^+_r)}$.  $N_{mn} ^{rs}$ and $N ^r_m$ are the standard Neumann
function coefficients.
Since the conformal map from a smooth Riemann surface to the Mandelstam
diagram is non-singular except at the vertices, we expect the usual
three string vertex to work for the general light-cone sigma models, as
well.  The one difference is the zero modes.  The oscillators have
time-dependent frequencies due to the $X^+$ dependence in the spacetime
fields, so the vertex depends on $\tau _0$ in a more complicated
fashion.  Also, the zero mode integral that produces the momentum
conservation delta function may be altered by the backgrounds.  The
oscillator terms represented by the Neumann coefficients should not
change, however, and the zero mode modifications may be determined
explicitly in many cases.
We will not check for contact terms.  This completes the
requirements for the first quantized string field theory.  Scattering
amplitudes may be computed using the usual LSZ techniques.  The
wavefunctionals may be second quantized to obtain the full closed string
field theory.  We leave the details for future work.

\section{Conclusion}

This paper has described the formulation of the manifestly ghost-free
light-cone gauge for the second order action in the Polyakov picture.
The action is that of a two-dimensional sigma model, giving a
bosonic string theory with spacetime metric, antisymmetric tensor,
dilaton and tachyon fields.  These fields must have a symmetry
generated by a null, covariantly constant spacetime vector in
order for the light-cone gauge to be fixed.  Also, the theory must be
Weyl invariant.  These two conditions are satisfied by a large class of
non-trivial critical string theories, including time-dependent
wave-like backgrounds.  The conditions for Weyl invariance have been
computed within the light-cone gauge, reproducing the usual beta
functions.  The calculation of the dilaton beta function and the
critical dimension is somewhat unusual because of the absence of
propagating ghosts.

These results confirm the notion that the light-cone gauge is only
sensible in critical string theories; i.e.\ those that could have been
quantized in the conformal gauge.  The absence of a Lorentz anomaly,
as in two and three dimensions, does not guarantee a consistent
light-cone theory.  Still there are many interesting sigma models
which may be quantized in the light-cone gauge.  Each has a
consistent, unitary string quantum mechanics and a relatively
simple string field theory.  These models include the exactly
solvable models discussed in section four.  They also include
more complex, wave-like backgrounds.

The requirement of a null symmetry is not extremely restrictive,
but it may seem strange and superfluous considering the many
excluded models that can be quantized in conformal gauge.  It
is interesting to note, however, that models with a flat direction
possess an extra $N=2$ supersymmetry involving the ghosts \cite{N2}.
The light-cone theories may just be the unitary, $N=2$ models
in the conformal gauge.  It is certainly an interesting class
of string models which is largely unexplored.

\vspace{.2in}
\noindent
{\bf Acknowledgements}

\vspace{.1in}
I would like to thank David Gross for stimulating discussions
throughout this project.  I also thank Eric D'Hoker and David Kutasov
for helpful conversations.

\vspace{.8in}

\renewcommand{\baselinestretch}{1.1}

\end{document}